\documentclass[a4paper,11pt]{article}
\usepackage{jheppub} 
\usepackage[T1]{fontenc}
\usepackage{multirow}
\usepackage{color}
\usepackage{ulem}
\usepackage{rotating}
\usepackage{booktabs}
\usepackage{dcolumn}
\usepackage{array} 
\usepackage{multirow}
\usepackage{lineno} 
\usepackage{siunitx}
\usepackage{amsmath}
\usepackage{bm}
\usepackage{mathrsfs}
\usepackage{indentfirst}

\DeclareSIUnit{\bmm}{\bm{m}}

\DeclareSIUnit{\clight}{\textnormal{\textit{c}}}
\DeclareSIUnit{\clight}{\textnormal{\textit{c}}}

\newcommand{\ar}{\rightarrow}

\newcolumntype{d}{D{.}{.}{-1}}
\newcolumntype{e}{D{.}{.}{8}}
\newcolumntype{f}{D{.}{.}{18}}
\newcolumntype{h}{D{.}{.}{13}}
\newcolumntype{g}{D{.}{.}{12}}

\title{\protect\boldmath  Measurement of Born cross sections of $e^+e^-\to\Xi^0\bar{\Xi}^0$ and search for charmonium(-like) states at $\sqrt{s}$ = 3.51-4.95~\textbf{GeV}}
\collaborationImg{\includegraphics[width=0.25\textwidth]{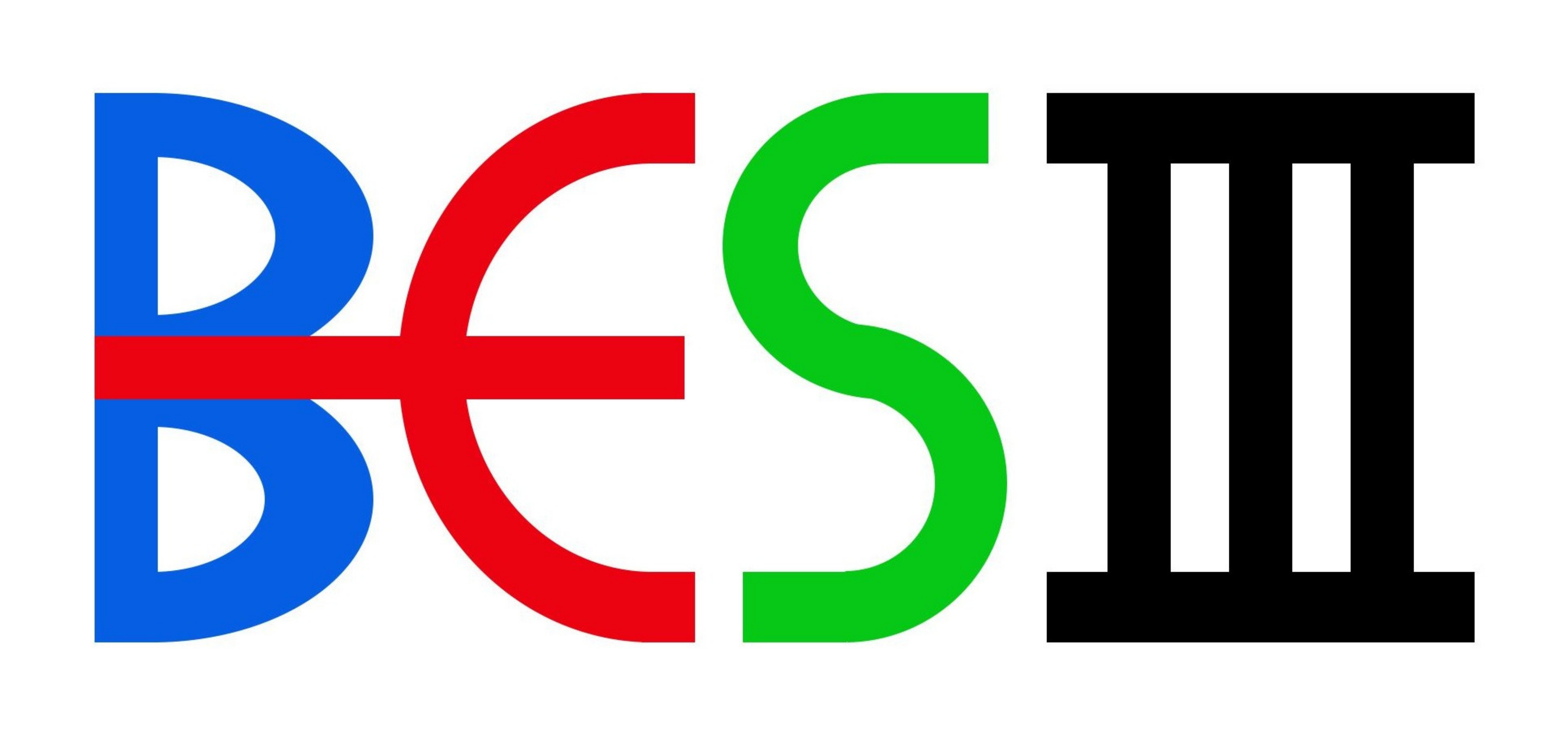}}
\collaboration{The BESIII collaboration}
\emailAdd{besiii-publications@ihep.ac.cn}

\begin{document} 
\abstract{Using $e^+e^-$ collision data collected by the BESIII
  detector at BEPCII corresponding to an integrated luminosity of 30
  $\rm fb^{-1}$, we measure Born cross sections and effective
  form factors for the process $e^+e^-\ar\Xi^0\bar{\Xi}^0$ at
  forty-five center-of-mass energies between 3.51 and 4.95~GeV.  The
  dressed cross section is fitted, assuming a power-law function plus
  a charmonium(-like) state, i.e., $\psi(3770)$, $\psi(4040)$,
  $\psi(4160)$, $\psi(4230)$, $\psi(4360)$, $\psi(4415)$ or
  $\psi(4660)$.  No significant charmonium(-like) state decaying into
  $\Xi^0\bar\Xi^0$ is observed.  Upper limits at the 90\% confidence
  level on the product of the branching fraction and the electronic
  partial width are provided for each decay.  In addition, ratios of
  the Born cross sections and the effective form factors for
  $e^+e^-\ar\Xi^0\bar{\Xi}^0$ and $e^+e^-\ar\Xi^-\bar{\Xi}^+$ are also
  presented to test isospin symmetry and the vector meson dominance
  model.

  }
\maketitle
\flushbottom    
\section{Introduction}
\label{sec:intro}
\noindent In 1974, the first member of charmonium family, the $J/\psi$
particle, was discovered~\cite{SLAC-SP-017:1974ind,E598:1974sol}.  It
provided support for the existence of a fourth quark, called the
charmed quark.  A series of charmonium states have been observed at
$e^+e^-$ colliders in the past decades. Three conventional charmonium
states, i.e., $\psi(4040)$, $\psi(4160)$, and
$\psi(4415)$~\cite{BES:2001ckj}, observed in the inclusive hadronic
cross section and dominated by open-charm final states, agree well
with the predictions of the potential model~\cite{Barnes:2005pb}.
Another five charmonium-like states, i.e. $\psi(4230)$, $\psi(4260)$,
$\psi(4360)$, $\psi(4634)$, and $\psi(4660)$, were observed via the initial
state radiation (ISR) process by the Belle and BaBar
experiments~\cite{BaBar:2005hhc,BaBar:2006ait,Belle:2007umv,Belle:2007dxy,Belle:2008xmh,BaBar:2012vyb,Belle:2013yex,BaBar:2012hpr,Belle:2014wyt},
and via direct production processes by the CLEO~\cite{CLEO:2006ike}
and BESIII experiments~\cite{BESIII:2014rja,BESIII:2023cmv}. The unexpected
multitude of states and mismatch of quantum numbers predicted by the
potential model have given rise to a great deal of interest.  Various
hypotheses have been proposed to explain their
nature~\cite{Yuan:2021wpg,Brambilla:2010cs,Briceno:2015rlt,Chen:2016qju,Close:2005iz,Wang:2019mhs,Qian:2021neg,Yan:2023yff},
including hybrid states, multiple-quark states, and molecular
structures.  However, up to now, no definitive conclusion has been
drawn.

This situation indicates an incomplete understanding of the strong
interaction, and to help clarify this, more experimental information
is needed. In particular, the study of two-body baryonic decays of
charmonium(-like) states, $\psi(3770)$, $\psi(4040)$,
  $\psi(4160)$, $\psi(4230)$, $\psi(4360)$, $\psi(4415)$ and
  $\psi(4660)$,
and their subsequent
hadronic decays in $e^+e^-$ collisions provide a new window for
understanding the nature of these states.  In addition, the
measurement of the electromagnetic form factors or the effective form
factor would also provide insight into the internal structure of the
charmonium(-like) states.  Although many experimental
studies~\cite{Ablikim:2013pgf, BESIII:2021ccp,Ablikim:2019kkp,BESIII:2023rse,BESIII:2021cvv,
  Wang:2022bzl,BESIII:2022kzc,BESIII:2023rwv, BESIII:2024umc,BESIII:2024ogz} of baryon anti-baryon
($B\bar{B}$) pair production have been performed above open-charm
threshold by the BESIII and Belle experiments, the only evidence for
$B\bar{B}$ final states from vector charmonium(-like)
decay is for $\psi(3770)\to\Lambda\bar{\Lambda}$ and
$\psi(3770)\to\Xi^-\bar{\Xi}^+$~\cite{BESIII:2021ccp,BESIII:2023rse}.
Thus, more precise measurements of exclusive cross sections for the
$e^+e^-\to B\bar{B}$ reactions are needed to further investigate the
nature of the charmonium(-like) states above open charm threshold.  Further, as
proposed by ref.~\cite{Iachello:1972nu}, the measured ratios of the
Born cross section or the effective form factor between the $e^+e^-\to
\Xi^0\bar{\Xi}^0$ process and its isospin partner processes is
important to validate the prediction based on the vector meson
dominance model~\cite{Iachello:2004aq,Bijker:2004yu,Yang:2019mzq,Li:2021lvs,Dai:2023vsw}.

In this article, we present measurements of the Born cross sections and
the effective form factors for the $e^+e^- \to \Xi^0\bar{\Xi}^0$
reaction, in the range of center-of-mass (CM) energy $\sqrt{s}$
between 3.51 and 4.95 GeV. These measurements are based on $e^+e^-$ collision data with a total integrated luminosity of 30 ${\rm fb}^{-1}$, collected by
the BESIII detector~\cite{besiii} at
BEPCII~\cite{BEPCII}. Measurements from the CLEO-c
experiment~\cite{Dobbs:2017hyd} are also shown for
comparison. Possible resonances are searched for by analyzing the line
shape of the dressed cross sections of the $e^+e^- \to
\Xi^0\bar{\Xi}^0$ reaction.  The product of branching fractions and
electronic partial widths for charmonium(-like) states
decaying into $\Xi^0\bar{\Xi}^0$ as well as their upper
limits at the 90\% confidence level (C.L.) are reported.  In addition,
the ratios of the Born cross section and the effective form factor for
the reactions of $e^+e^-\to\Xi^0\bar{\Xi}^0$ and
$e^+e^-\to\Xi^-\bar{\Xi}^+$ from the BESIII
experiment~\cite{BESIII:2023rse} are presented.

\section{BESIII Detector and Monte Carlo simulation}
\noindent The BESIII detector~\cite{besiii} records symmetric $e^+e^-$
collisions provided by the BEPCII storage ring~\cite{BEPCII} in the
range of $\sqrt{s}$ from 1.85 to \SI{4.95}{GeV}, with a peak
luminosity of \SI{1.1e33} {\per\centi\meter\squared\per\second}
achieved at $\sqrt{s} =$ 3.773 {GeV}. BESIII has collected large data
samples in this energy region~\cite{Ablikim:2019hff, EcmsMea,
  EventFilter}. The cylindrical core of the BESIII detector covers
93\% of the full solid angle and consists of a helium-based multilayer
drift chamber~(MDC), a time-of-flight system~(TOF), and a CsI(Tl)
electromagnetic calorimeter~(EMC), which are all enclosed in a
superconducting solenoidal magnet providing a \SI{1.0}{T} magnetic
field. The solenoid is supported by an octagonal flux-return yoke with
resistive plate counter muon identification modules interleaved with
steel. The charged-particle momentum resolution at \SI{1}{GeV/\clight}
is $0.5\%$, and the ${\rm d}E/{\rm d}x$ resolution is $6\%$ for
electrons from Bhabha scattering. The EMC measures photon energies
with a resolution of $2.5\%$ ($5\%$) at \SI{1}{GeV} in the barrel (end
cap) region. The time resolution of the plastic-scintillator TOF
system in the barrel region is \SI{68}{ps}, while that in the end
cap region was \SI{110}{ps}. The end cap TOF system was upgraded in
2015 using multigap resistive plate chamber technology, providing a
time resolution of \SI{60}{ps}~\cite{etof1,etof2,etof3} and benefiting
82\% of the data used in this analysis.

To determine the detection efficiency, simulated samples of
$4\times10^5$ $e^+e^-\to\Xi^0\bar{\Xi}^0$ events are produced for each
of the forty-five CM energy points from 3.51 to 4.95 GeV with {\sc
  geant4}-based~\cite{GEANT4} Monte Carlo (MC) software, which
includes the geometric description of the BESIII detector and the
detector response~\cite{Huang:2022wuo}. The simulation models the beam
energy spread and initial state radiation (ISR) in the $e^+e^-$
annihilations with the generator {\sc kkmc}~\cite{KKMC}. The
$e^+e^-\to\Xi^0\bar{\Xi}^0$ process and its subsequent decays are
simulated with a uniform phase space (PHSP) model by {\sc
  evtgen}~\cite{evtgen2,EVTGEN}.

\section{Event selection}
\noindent Reconstructing $e^+e^-\to\Xi^0\bar{\Xi}^0$ candidate events
fully has low efficiency. To achieve higher efficiency, a single
baryon $\Xi^0$ tag technique is employed, i.e., only the $\Xi^0$
baryon is reconstructed via its decay $\Xi^0\to\pi^0\Lambda$ with the subsequent decays $\Lambda\to p\pi^-$ and $\pi^0\to\gamma\gamma$, and the antibaryon $\bar\Xi^0$
is identified in the $\Xi^0$ recoil mass distribution. The charge conjugate decays are implied unless otherwise noted.

Charged tracks are required to be within the angular coverage of the
MDC: $|\rm{cos}\theta|< 0.93$, where $\theta$ is the polar angle with
respect to the $z$-axis, which is the symmetry axis of the MDC. At
least one positive and one negative charged track well reconstructed
in the MDC are required.

Particle identification (PID) for charged tracks combines measurements
of the specific ionization energy loss in the MDC ($dE/dx$) and the
flight time in the TOF to form likelihoods $\mathcal{L}(h)$ ($h = p, K,
\pi$) for each hadron $h$ hypothesis. Tracks are identified as
protons when the proton hypothesis has the greatest likelihood
($\mathcal{L}(p) > \mathcal{L}(K)$ and $\mathcal{L}(p) > \mathcal{L}(\pi)$), while charged
pions are identified when $\mathcal{L}(\pi) > \mathcal{L}(K)$ and $\mathcal{L}(\pi) > \mathcal{L}(p)$.  Events with at least one $p$ and one $\pi^-$ are kept for further analysis.

Photons are reconstructed from isolated showers in the EMC.  The
energy deposited in the nearby TOF counter is included to improve the
reconstruction efficiency and energy resolution.  The energies of
photons are required to be greater than 25 MeV in the EMC barrel
region ($|\cos\theta|<0.8$), and greater than 50 MeV in the EMC
end-cap region ($0.86<|\cos\theta|<0.92$).  Furthermore, to suppress electronic noise and showers unrelated to the event, the difference between the EMC time and the event start time is required to be within 
[0, 700]\,ns. Events with at least two
photons are kept for further analysis.

To reconstruct the $\pi^0$ meson from the $\Xi^0\to\pi^0\Lambda$
decay, a one-constraint (1C) kinematic fit is applied to all
$\gamma\gamma$ combinations under the hypothesis of
$\pi^0\to\gamma\gamma$, constraining the invariant mass of two photons
to the $\pi^0$ mass, and $\chi^{2}_{1\rm C}$ $\leq 20$ is required to
suppress the non-$\pi^0$ background~\cite{BESIII:2016nix,
  BESIII:2019dve, BESIII:2021aer}.

To reconstruct the $\Lambda$ candidate, a secondary vertex fit looping
over all $p\pi^-$ combinations is employed, and the corresponding
$\chi^2$ value is required to be less than 500 by default. The distance between the interaction point and the decay
vertex of the $\Lambda$ candidate must be greater than zero. Further, a
requirement of $|M_{p\pi^-} - m_{\Lambda}| \leq$ 5 MeV/${c^2}$ is
imposed, where $M_{p\pi^-}$ is the invariant mass of the $p\pi^-$
combination and $m_{\Lambda}$ is the $\Lambda$ mass~\cite{PDG2022}.
The requirement is determined by the figure-of-merit (FOM =
${\cal{S}}^{\prime}/\sqrt{{\cal{S}}^{\prime} + B}$), where
${\cal{S}}^{\prime}$ is the number of signal MC events and $B$ is the
number of the estimated background events.

\begin{figure}[h] \centering
\includegraphics[width=0.8\textwidth]{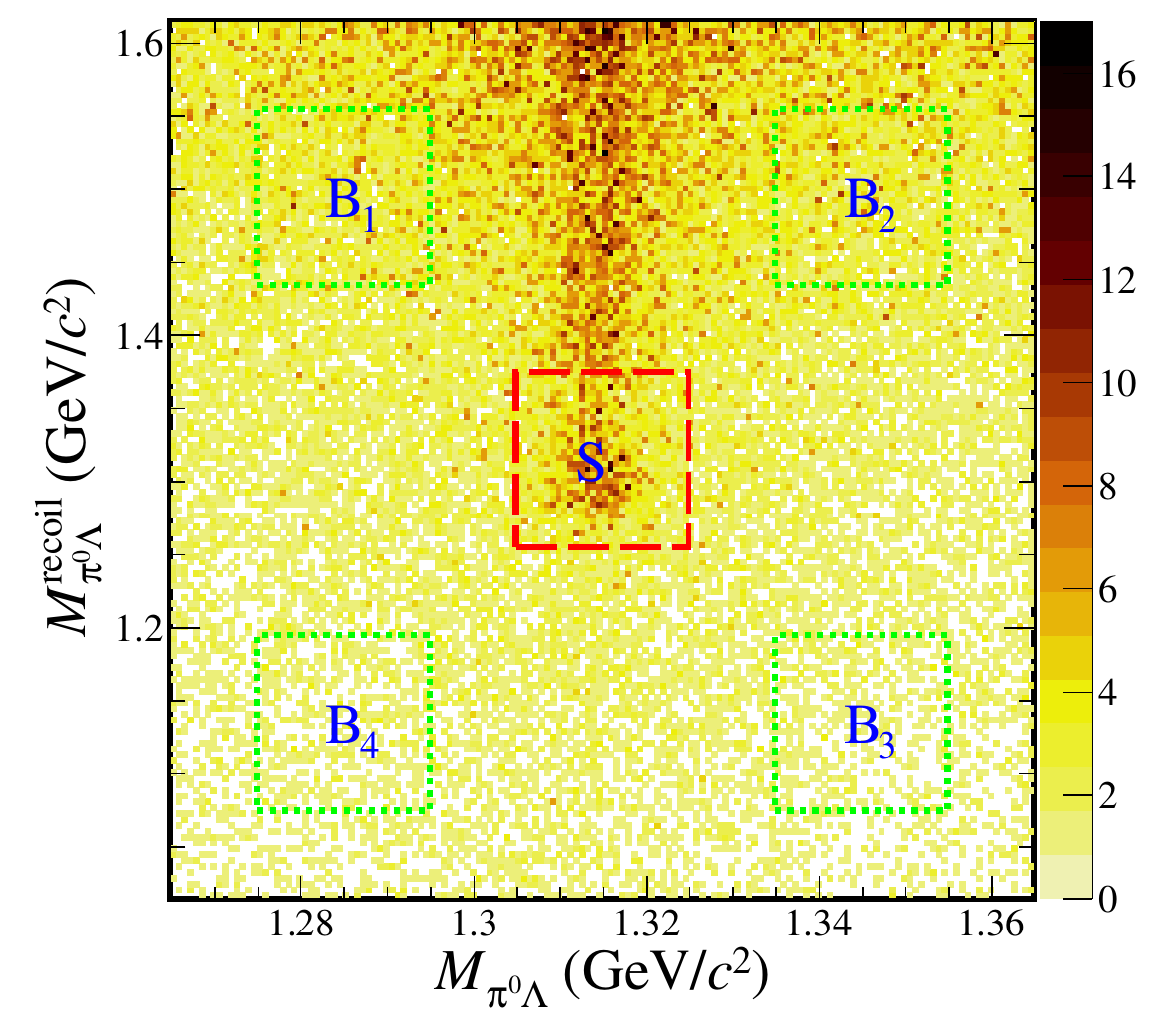}
\caption{\small {The distribution of $M_{\pi^0\Lambda}^{\rm recoil}$
    versus $M_{\pi^0\Lambda}$ of the accepted candidates of data
    summed over CM energy points. The red box represents the signal
    region, and the green boxes represent sideband regions.}}
    \label{eachdata}
\end{figure}

The $\Xi^0$ candidate with the minimum value of $|M_{\pi^0\Lambda} -
m_{\Xi^0}|$ of all $\pi^0\Lambda$ combinations is selected, where
$M_{\pi^0\Lambda}$ is the invariant mass of the $\pi^0\Lambda$ system,
and $m_{\Xi^0}$ is the $\Xi^0$ mass~\cite{PDG2022}.
The mass recoiling against the selected $\pi^0\Lambda$ system is given
by \begin{equation} M^{\rm recoil}_{\pi^{0}\Lambda} = \sqrt{(\sqrt{s}
-E_{\pi^{0}\Lambda})^{2} - |\vec{p}_{\pi^{0}\Lambda}|^2},
\end{equation} where $E_{\pi^0\Lambda}$ and $\vec{p}_{\pi^0\Lambda}$
are the energy and momentum, respectively, of the selected
$\pi^0\Lambda$ candidate in the CM system. The anti-baryon
$\bar{\Xi}^0$ candidate is required to be in the signal mass window
$|M_{\pi^0\Lambda}- m_{\Xi^0}| \leq$ 10 MeV/${c^2}$ and $|M^{\rm
recoil}_{\pi^0\Lambda}- m_{\Xi^0}| \leq$ 60 MeV/${c^2}$, marked by S
as shown in figure~\ref{eachdata}.  A clear accumulation around the
$\Xi^0$ mass can be seen in figure~\ref{eachdata}. The
    $M^{\rm recoil}_{\pi^0\Lambda}$ tail from the $\gamma\Xi^0
    \bar{\Xi}^0$, $\gamma\gamma\Xi^0 \bar{\Xi}^0$, and
    $\gamma\gamma\gamma\Xi^0 \bar{\Xi}^0$ processes contribute
    negligibly to the signal yields.

\section{Born cross section measurement}
\subsection{Determination of signal yields}

\noindent After applying the above requirements, the remaining
background events mainly come from the $e^+e^-\to
\pi^0\pi^0\Lambda\Lambda$ process with the same final state as the
signal channel. The background yields in the signal region are
evaluated using four sideband regions $B_i$, where $i$ runs over 1 to
4, each with the same area as the signal regions.  The sideband
regions are defined in the mass windows of $M_{\pi^0\Lambda}$ and
$M^{\rm recoil}_{\pi^0\Lambda}$ as shown in figure~\ref{eachdata}, i.e.,
$B_{1}$: [1.275, 1.295] and [1.435, 1.555] GeV/$c^2$, $B_{2}$: [1.335,
1.355] and [1.435, 1.555] GeV/$c^2$, $B_{3}$: [1.335, 1.355] and
[1.075, 1.195] GeV/$c^2$, $B_{4}$: [1.275, 1.295] and [1.075, 1.195]
GeV/$c^2$.  The signal yield $N_{\rm obs}$ for the
$e^+e^-\ar\Xi^0\bar{\Xi}^0$ process at each CM energy point is then
determined by $N_{\rm obs} = N_{\rm S} - N_{\rm
bkg}$, where $N_S$ represents the number of events in the signal
region and $N_{\rm bkg}$ is the number of background events,
i.e., $N_{\rm bkg} = \frac{1}{4}\sum^{4}_{i=1} N_{B_{i}}$. The
uncertainty of $N_{\rm obs}$ and its upper limit are computed by the
TRolke method~\cite{Lundberg:2009iu}. The numerical results are summarized in
table~\ref{tab:signal:yields:DD}.

\subsection{Determination of Born cross section and effective form
factor} 
\begin{figure}[h] \centering
\includegraphics[width=0.99\textwidth]{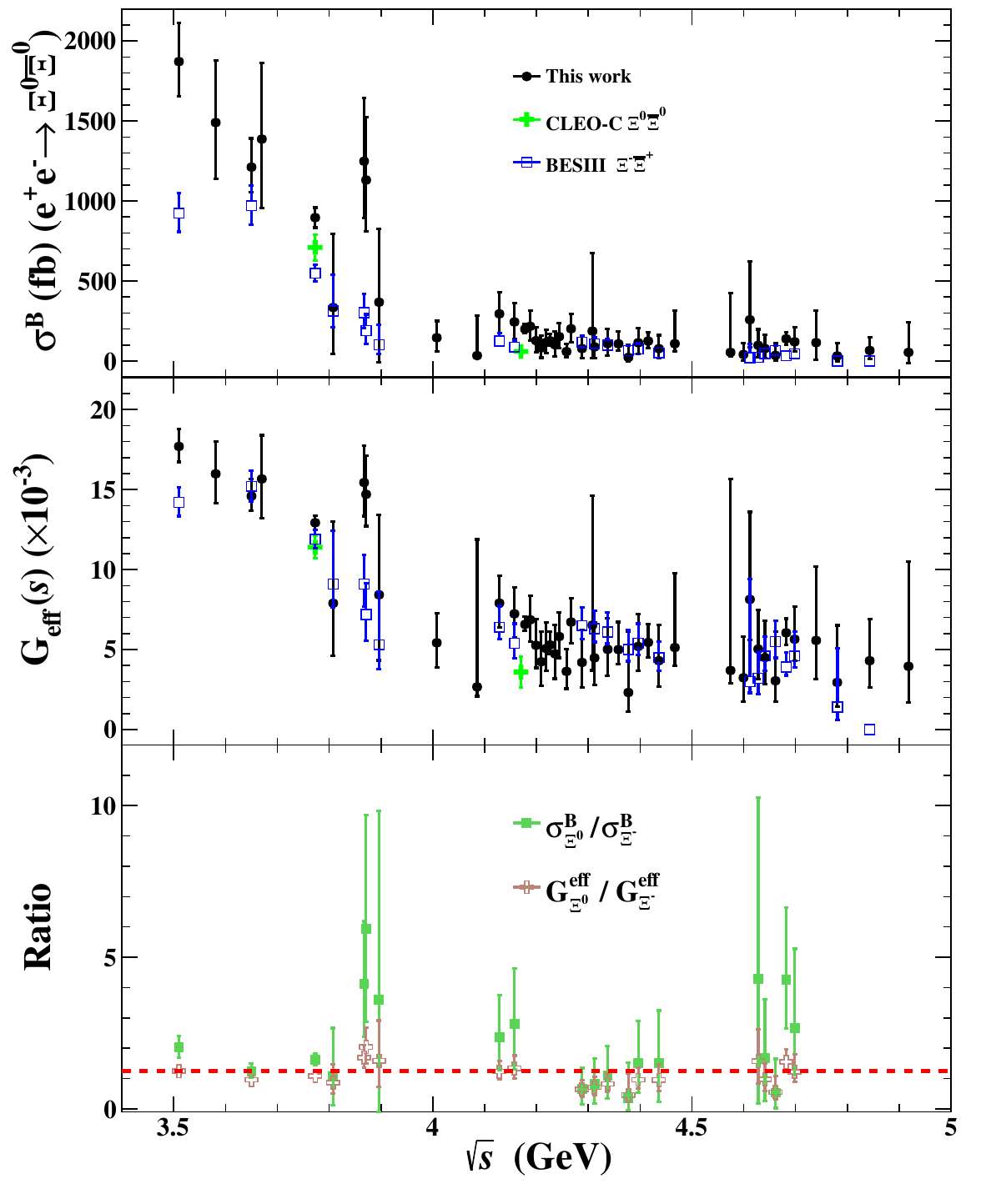}
\caption{Comparisons of the Born cross section ($\sigma^{B}$) and the
  effective form factor ($G_{\rm eff}(s)$) for
  $e^+e^-\to\Xi^0\bar{\Xi}^0$ from this work and
  CLEO-c~\cite{Dobbs:2017hyd} with $e^+e^-\to\Xi^-\bar{\Xi}^+$ from
  BESIII \cite{BESIII:2023rse} as a function of CM energy. The bottom
  one shows the ratios of the Born cross sections and the effective
  form factors for the reactions of $e^+e^-\to\Xi^0\bar{\Xi}^0$ and
  $e^+e^-\to\Xi^-\bar{\Xi}^+$, the red dashed line represents the
  theoretically predicted cross section ratio~\cite{Dai:2023vsw}. Here
  the uncertainties include both systematic and statistical
  uncertainties.}  \label{Fig:ratio_of_sig} \end{figure}

\noindent The Born cross section of the
$e^+e^-\to\Xi^0\bar{\Xi}^0$ process is given by \begin{equation}
  \sigma^{B} =\frac{N_{\rm obs}}{2{\cal{L}}(1 + \delta)\frac{1}{|1 -
      \Pi|^{2}}\epsilon\cal{B}}, \end{equation} where the factor of 2
accounts for the charge conjugate mode being included, ${\cal{L}}$ is
the integrated luminosity, $(1 + \delta)$ is the ISR correction
factor, $\frac{1}{|1-\Pi|^2}$ is the vacuum polarization (VP)
correction factor, $\epsilon$ is the detection efficiency, and ${\cal
  B}$ represents the products of the branching fractions of
$\Xi^0\to\pi^0\Lambda$, $\Lambda\to p\pi^-$ and
$\pi^0\to \gamma\gamma$~\cite{PDG2022}. The VP correction factor is
calculated according to ref.~\cite{Jegerlehner:2011ti}. The results of
the measured Born cross section for each CM energy point are listed in
table~\ref{tab:signal:yields:DD}. Note that the single-baryon tag
method causes double counting of the $\Xi^0\bar{\Xi}^0$ final
state. To correct for this effect, a factor of 1.08 is taken into
account when calculating the statistical uncertainties based on the
study of MC simulation~\cite{BESIII:2021aer}. The efficiency and ISR
correction factor are obtained through an iterative
process~\cite{Sun:2020ehv}.  The measured Born cross section at each
CM energy point is shown in figure~\ref{Fig:ratio_of_sig} together
with the results from the charged mode $e^+e^-\to\Xi^-\bar{\Xi}^+$
from the BESIII~\cite{BESIII:2023rse} and $\Xi^0\bar{\Xi}^0$ results
from CLEO-c~\cite{Dobbs:2017hyd} measurements. Our results are roughly
consistent with those of CLEO-c at $\sqrt{s} = 3.770$ and 4.160 GeV.
Figure~\ref{Fig:ratio_of_sig} also shows the energy dependence of the
$\Xi^0$ effective form factor $G_{\rm eff}(s)$ compared with the
CLEO-c results at $\sqrt{s} = 3.770$ and 4.160
GeV~\cite{Dobbs:2017hyd}. $|G_{\rm eff}(s)|$ is defined
as~\cite{Ablikim:2019kkp} \begin{equation} |G_{\rm eff}(s)| =
  \sqrt{\frac{3s\sigma^B}{4 \pi\alpha^2\beta(\frac{1}{2\tau}+1)}},
\end{equation}
where $\alpha = \frac{1}{137}$ is the fine structure constant, $\beta
= \sqrt{1-\frac{1}{\tau}}$ is the velocity of $\Xi^{0}$ in CM system, $\tau =
\frac{s}{4m_{\Xi^0}^2}$, and $m_{\Xi^0}$ is the $\Xi^0$ 
mass~\cite{PDG2022}. Also shown in figure~\ref{Fig:ratio_of_sig} are the
ratios of Born cross sections and the effective form factors for the
reactions of $e^+e^-\to\Xi^0\bar{\Xi}^0$ and $e^+e^- \to$
$\Xi^-\bar{\Xi}^+$ from the BESIII experiment~\cite{BESIII:2023rse}.

\begin{table}[!htbp] \begin{center} \caption{ The Born cross section
$\sigma^{B}$ and the effective form factor $|G_{\rm eff}(s)|$ for
$e^+e^- \to \Xi^0\bar{\Xi}^0$ at forty-five energy points between 3.51
and 4.95 GeV. The values in the brackets are the corresponding upper
limits at the 90\% C.L.. The first uncertainties are statistical, and the
second are systematic. $\sqrt{s}$ is the
$e^{+}e^{-}$ CM energy~\cite{BESIII:2015zbz, BESIII:2020eyu},
$\mathcal{L}$ is the integrated luminosity of each data
set~\cite{BESIII:2015qfd,BESIII:2022dxl,BESIII:2022ulv}, $\frac{1}{|1-\Pi|^2}$ is the vacuum
polarization correction factor, and $\epsilon
(1+\delta)$ is the product of
the ISR correction factor and the detection efficiency. $N_{\rm S}$ is the number of events in the signal region, 
$N_{\rm bkg}$ is the number of background events scaled from the
sideband region, and $N_{\rm obs}$ is the number of observed events
after subtracting $N_{\rm bkg}$ from $N_{\rm S}$ with the
uncertainty calculated by the TRolke method~\cite{Lundberg:2009iu}
(the number of signal events for the upper limit with the
consideration of systematic uncertainty estimated based on the TRolke
method~\cite{Lundberg:2009iu}). $\sigma^B$ is the Born cross
section, $|G_{\rm eff}(s)|$ is the effective form factor, and $S$ is
the statistical significance.}
    \resizebox{0.95\textwidth}{!}{%
    \begin{tabular}{l c c c c c  l l@{}l l c}
        \hline
        \hline
		$\sqrt{s}$ (GeV) &$\mathcal{L}\rm~(pb^{-1})$ & $\frac{1}{|1-\Pi|^2}$ & $\epsilon (1+\delta)$  &$N_{\rm S}$ &$N_{\rm bkg}$ &~~~~$N_{\rm obs}$     &~~~~$\sigma^{B}$ (fb)&$|G_{\rm eff}(s)| ~(\times 10^{-3})$ &$S(\sigma)$ \\
        \hline
3.51000  &405.7 &1.04 &12.8 &212	&83.8 &128.3$^{+12.6}_{-10.6}$    &1871$^{+199}_{-167}$ $\pm120$  &17.7$^{+0.9}_{-0.8}$ $\pm0.6$    &11.7\\
3.58100   &85.7  &1.04 &13.0 &39	&17.3 &21.8$^{+4.9}_{-4.4}$      &1490$^{+363}_{-326}$ $\pm96$  &16.0$^{+1.9}_{-1.7}$ $\pm0.5$  &4.6\\
3.65000   &410.0 &1.02 &12.7 &126	&44.8 &81.3$^{+9.8}_{-8.3}$     &1212$^{+158}_{-134}$ $\pm78$  &14.6$^{+0.9}_{-0.8}$ $\pm0.5$   &9.9\\
3.67000   &84.7  &0.99 &12.5 &26	    &7.8 &18.3$^{+5.8}_{-5.2}$     &1387$^{+476}_{-427}$ $\pm89$  &15.6$^{+2.7}_{-2.4}$ $\pm0.5$    &5.1\\
3.77300   &7926.8&1.06 &12.2 &2162	&1004 &1158.0$^{+35.0}_{-34.0}$    &896$^{+29}_{-28}$  $\pm56$   &12.8$^{+0.2}_{-0.2}$ $\pm0.4$ &8.1\\
3.80765 &50.5  &1.06 &12.2 &8	    &5.3 &2.8$^{+3.3}_{-2.1}$ ($< 9.5$)      &334$^{+433}_{-276}$ $\pm22 $ $(<1159)$   &7.9$^{+5.1}_{-3.3}$ $\pm0.2$ $(<14.7)$  &1.4\\
3.86741 &108.9 &1.05 &11.8 &29	    &7.8 &21.3$^{+5.8}_{-5.2}$       &1249$^{+368}_{-331}$ $\pm80$  &15.4$^{+2.3}_{-2.0}$ $\pm0.5$ &5.9\\
3.87131 &110.3 &1.05 &11.7 &25	    &5.8 &19.3$^{+5.8}_{-4.7}$      &1131$^{+368}_{-298}$ $\pm73$  &14.7$^{+2.4}_{-1.9}$ $\pm0.5$ &5.9\\
3.89624 &52.6  &1.05 &11.7 &7	    &4.0 &3.0$^{+3.3}_{-2.7}$ ($< 9.1$)        &368$^{+437}_{-357}$ $\pm24$ $(<1119)$   &8.4$^{+5.0}_{-4.1}$ $\pm0.3$ $(<14.7)$ &1.6\\
4.00762 &482.0 &1.04 &10.8 &30 	&20.0 &10.0$^{+6.3}_{-5.2}$ ($< 22.3$)          &146$^{+99}_{-82}$ $\pm9$ $(<326)$   &5.4$^{+1.8}_{-1.5}$ $\pm0.2$ $(<8.1)$ &2.3\\
4.08545 &52.9  &1.05 &10.4 &2	    &1.8 &0.3$^{+1.6}_{-0.1}$ ($< 4.6$)         &34$^{+237}_{-15}$ $\pm2$ $(<632)$ &2.7$^{+9.2}_{-0.6}$ $\pm0.1$  $(<11.5)$ &0.6\\
4.12848 &401.5 &1.05 &9.4 &26	    &11.3 &14.8$^{+5.8}_{-5.2}$       &295$^{+125}_{-112}$ $\pm19$  &7.9$^{+1.7}_{-1.5}$ $\pm0.3$  &3.8\\
4.15744 &408.7 &1.05 &9.4 &22	    &9.5 &12.5$^{+5.3}_{-4.7}$       &244$^{+112}_{-99}$ $\pm16$   &7.2$^{+1.7}_{-1.5}$ $\pm0.2$   &3.6\\
4.17800   &3189  &1.05 &9.7 &154	&72.3 &81.8$^{+9.8}_{-8.3}$        &200$^{+26}_{-22}$ $\pm13$   &6.6$^{+0.4}_{-0.4}$ $\pm0.2$ &8.2\\
4.18800   &526.7 &1.05 &9.5 &27	    &12.5 &14.5$^{+5.8}_{-5.2}$         &217$^{+94}_{-84}$ $\pm14$     &6.9$^{+1.5}_{-1.3}$ $\pm0.2$   &3.7\\
4.19915 &526.0 &1.06 &9.5 &17	    &8.5 &8.5$^{+4.8}_{-4.2}$ ($< 18.0$)       &127$^{+78}_{-68}$ $\pm8$ $(<271)$   &5.3$^{+1.6}_{-1.4}$ $\pm0.2$  $(<7.7)$ &2.7\\
4.20939  &517.1 &1.06 &9.3 &15   &9.8 &5.3$^{+4.3}_{-3.4}$ ($< 14.7$)          &82$^{+73}_{-57}$ $\pm5$ $(<231)$   &4.2$^{+1.9}_{-1.5}$ $\pm0.1$  $(<7.1)$ &1.8\\
4.21893  &514.6 &1.06 &9.0 &15	    &7.8 &7.3$^{+4.3}_{-3.7}$ ($< 16.5$)         &117$^{+75}_{-65}$ $\pm8$ $(<267)$     &5.1$^{+1.6}_{-1.4}$ $\pm0.2$  $(<7.7)$ &2.5\\
4.22626 &1100.9&1.06 &9.4 &40	    &22.5 &17.5$^{+4.9}_{-3.5}$      &127$^{+39}_{-28}$ $\pm8$    &5.3$^{+0.8}_{-0.6}$ $\pm0.2$   &3.5\\
4.23570  &530.3 &1.06 &9.4 &18	    &11.3 &6.8$^{+4.8}_{-4.1}$ ($< 16.5$)        &101$^{+78}_{-67}$ $\pm7$ $(<249)$   &4.7$^{+1.8}_{-1.6}$ $\pm0.2$  $(<7.4)$ &2.1 \\
4.24397  &538.1 &1.06 &9.2 &19	    &9.0 &10.0$^{+4.8}_{-4.2}$        &153$^{+79}_{-69}$ $\pm10$    &5.8$^{+1.5}_{-1.3}$ $\pm0.2$  &3.0\\
4.25797 &828.4 &1.05 &9.2 &20	    &14.0 &6.0$^{+4.3}_{-3.3}$ ($< 16.1$)          &59  $^{+46}_{-35}$ $\pm4$ $(<160)$    &3.6$^{+1.4}_{-1.1}$ $\pm0.1$  $(<6.0)$ &1.8\\
4.26681  &531.1 &1.05 &9.1 &21	    &8.0 &13.0$^{+5.3}_{-4.7}$           &202$^{+89}_{-79}$ $\pm13$    &6.7$^{+1.5}_{-1.3}$ $\pm0.2$   &3.9\\
4.28788 &502.4 &1.05 &8.2 &11	    &6.8 &4.3$^{+4.1}_{-3.1}$ ($< 12.4$)       &78$^{+76}_{-58}$ $\pm5$ $(<229)$         &4.2$^{+2.0}_{-1.5}$ $\pm0.1$  $(<7.2)$ &1.7\\
4.30789 &45.1  &1.05 &8.9 &2	    &1.0 &1.0$^{+2.3}_{-0.9}$ ($< 4.6$)           &187$^{+465}_{-162}$ $\pm12$ $(<864)$ &6.5$^{+8.1}_{-2.8}$ $\pm0.2$  $(<14.0)$ &1.1\\
4.31205 &501.2 &1.05 &8.1 &13	    &8.3  &4.8$^{+4.3}_{-3.3}$ ($< 13.1$)        &88$^{+86}_{-66}$ $\pm6$ $(<244)$        &4.5$^{+2.2}_{-1.7}$ $\pm0.1$  $(<7.5)$ &1.8\\
4.33739 &505.0 &1.05 &8.2 &14	    &8.0 &6.0$^{+4.6}_{-3.6}$ ($< 14.3$)         &109$^{+85}_{-71}$ $\pm7$ $(<261)$      &5.0$^{+1.9}_{-1.6}$ $\pm0.2$  $(<7.8)$    &2.1\\
4.35826 &543.9 &1.05 &8.7 &12	    &5.3 &6.8$^{+4.3}_{-2.2}$ ($< 14.5$)         &108$^{+74}_{-38}$  $\pm7$ $(<232)$     &5.0$^{+1.7}_{-0.9}$ $\pm0.2$  $(<7.4)$ &2.7\\
4.37737 &522.7 &1.05 &7.9 &11	    &9.8 &1.3$^{+3.8}_{-1.2}$ ($< 9.8$)       &23$^{+75}_{-24}$ $\pm1$ $(<181)$        &2.3$^{+3.8}_{-1.2}$ $\pm0.1$  $(<7.8)$ &0.9\\
4.39645 &507.8 &1.05 &7.8 &12	    &6.0 &6.0$^{+4.3}_{-3.2}$ ($< 13.7$)          &115$^{+89}_{-66}$ $\pm7$ $(<262)$      &5.2$^{+2.0}_{-1.5}$ $\pm0.2$  $(<7.9)$ &2.3\\
4.41558 &1090.7&1.05 &8.2 &29	    &14.3 &14.8$^{+5.8}_{-4.7}$                 &125$^{+53}_{-43}$ $\pm8$    &5.4$^{+1.2}_{-0.9}$ $\pm0.2$   &3.5\\
4.43624 &569.9 &1.05 &7.7 &15	    &10.5 &4.5$^{+4.3}_{-3.1}$ ($< 13.7$)       &77$^{+80}_{-58}$ $\pm5$ $(<238)$        &4.3$^{+2.2}_{-1.6}$ $\pm0.2$  $(<7.5)$ &1.6\\
4.46706 &111.1 &1.05 &7.8 &2	    &0.8 &1.3$^{+2.1}_{-0.5}$ ($< 5.4$)        &108$^{+197}_{-47}$ $\pm7$ $(<470)$   &5.1$^{+4.6}_{-1.1}$ $\pm0.2$  $(<18.0)$ &1.4\\
4.57450 &48.9  &1.05 &7.1 &1	    &0.8 &0.3$^{+1.5}_{-0.1}$ ($< 3.7$)        &54$^{+351}_{-23} $ $\pm3$ $(<803)$     &3.7$^{+12.0}_{-0.8}$ $\pm0.1$  $(<14.2)$ &0.6\\
4.59953 &586.9 &1.06 &7.0 &8	    &5.8 &2.3$^{+3.3}_{-1.9}$ ($< 9.5$)         &41$^{+65}_{-38}$ $\pm3$ $(<174)$       &3.2$^{+2.6}_{-1.5}$ $\pm0.1$  $(<6.7)$ &1.2\\
4.61186 &103.7 &1.05 &6.3 &4      &1.8 &2.3$^{+2.8}_{-1.3}$ ($< 7.5$)       &258$^{+347}_{-162}$ $\pm17$ $(<863)$  &8.1$^{+5.5}_{-2.5}$ $\pm0.3$  $(<15.0)$ &1.6\\
4.62800 &521.5 &1.05 &6.2 &11	    &6.8 &4.3$^{+3.8}_{-2.9}$ ($< 12.4$)        &98$^{+95}_{-72}$ $\pm6$ $(<287)$     &5.0$^{+2.4}_{-1.9}$ $\pm0.2$  $(<8.6)$ &1.7\\
4.64091 &551.7 &1.05 &6.1 &8	    &4.5 &3.5$^{+3.3}_{-2.4}$ ($< 10.4$)       &78$^{+79}_{-58}$ $\pm5$ $(<233)$      &4.5$^{+2.3}_{-1.7}$ $\pm0.1$  $(<7.8)$ &1.7\\
4.66124 &529.4 &1.05 &6.0 &4	    &2.5 &1.5$^{+2.8}_{-1.2}$ ($< 6.7$)        &36$^{+72}_{-31}$ $\pm2$ $(<159)$      &3.0$^{+3.1}_{-1.3}$ $\pm0.1$  $(<6.4)$ &1.2\\
4.68192 &1667.4 &1.05 &5.9 &25	    &6.3  &18.3$^{+4.8}_{-4.2}$       &139$^{+39}_{-34}$ $\pm9$     &6.0$^{+0.9}_{-0.8}$ $\pm0.2$ &5.7\\
4.69822 &535.5 &1.06 &5.9 &6  	&1.0 &5.0$^{+3.3}_{-2.1}$         &120$^{+86}_{-57} $ $\pm8$     &5.6$^{+2.0}_{-1.3}$ $\pm0.2$   &3.4\\
4.73970 &163.9  &1.06 &5.9 &3	    &1.5 &1.5$^{+2.3}_{-1.2}$ ($< 6.1$)   &115$^{+191}_{-100}$ $\pm7$ $(<470)$ &5.6$^{+4.6}_{-2.4}$ $\pm0.2$  $(<11.2)$ &1.3\\
4.78054 &511.5 &1.06 &5.7 &5	    &3.8 &1.3$^{+2.3}_{-1.2}$ ($< 7.2$)       &32$^{+77}_{-33}$ $\pm2$ $(<184)$  &2.9$^{+3.6}_{-1.5}$ $\pm0.1$  $(<7.1)$ &1.0\\
4.84307 &525.2 &1.06 &5.4 &5	    &2.5 &2.5$^{+2.8}_{-1.8}$ ($< 8.1$)       &66$^{+80}_{-51}$ $\pm4$ $(<215)$      &4.3$^{+2.6}_{-1.7}$ $\pm0.1$  $(<7.7)$ &1.6\\
4.91802 &207.8 &1.06 &5.0 &3	    &2.3 &0.8$^{+2.3}_{-0.8}$ ($< 5.3$)         &54$^{+180}_{-62}$ $\pm3$ $(<384)$   &3.9$^{+6.5}_{-2.3}$ $\pm0.1$  $(<10.5)$ &0.8\\
        \hline
        \hline
    \end{tabular}}
    \label{tab:signal:yields:DD}
    \end{center}
\end{table}

\section{Systematic uncertainty} \noindent Systematic uncertainties on
the Born cross section measurements mainly originate from the
integrated luminosity, the $\Xi^0$ reconstruction, background, angular
distribution, branching fractions, and input line shape.
\subsection{Luminosity} 
\noindent The luminosity at each CM energy
point is measured using Bhabha events, with an uncertainty of
1.0\%~\cite{BESIII:2015qfd}, which is taken as the systematic
uncertainty due to the luminosity measurement.

\subsection{$\Xi^0$ reconstruction}

\noindent The systematic uncertainty due to the $\Xi^{0}$
reconstruction efficiency, incorporating the tracking and PID
efficiencies, the photon selection efficiency, the reconstruction
efficiencies of $\Lambda$ and $\pi^0$, the $\Lambda$ decay lengths,
and the mass windows of $\Lambda$ and $\Xi^{0}$, is evaluated by a
control sample of $\psi(3686)\rightarrow\Xi^0\bar\Xi^0$ with the same
method as used in refs.~\cite{BESIII:2012ghz,BESIII:2016ssr,BESIII:2021gca,BESIII:2020ktn,BESIII:2022mfx,BESIII:2022lsz,BESIII:2023lkg,BESIII:2023euh,BESIII:2024dmr,BESIII:2024jgy}.  The efficiency difference between data and MC
simulation is found to be 4.5\%, which is taken as the systematic
uncertainty of the $\Xi^0$ reconstruction.

\subsection{Background}

\noindent The systematic uncertainty due to the background is
estimated by shifting the sideband regions inward and outward by 10 MeV for $M_{\pi^0\Lambda}$ and 60 MeV for $M^{\text{recoil}}_{\pi^0\Lambda}$ from the signal region with a standard of 3$\sigma$ mass resolution, using the sum of all energy points. The resulting
largest difference of 1.2\% is taken as the systematic uncertainty due
to the background.

\subsection{Angular distribution}

\noindent The uncertainty due to the angular distribution is estimated
to be 3.5\% by comparing the efficiency of the angular distribution
from the phase space simulation with that incorporating the $\Xi^0$
transverse polarization and the spin correlation.

\subsection{Branching fractions} 

\noindent The uncertainty for the product of the branching fractions
of $\Lambda\to p\pi^-$ and $\pi^0\to\gamma\gamma$ is 0.8\% taken from
the Particle Data Group (PDG)~\cite{PDG2022}.

\subsection{Input line shape}

\noindent The uncertainty due to the input line shape of the cross
section for determining the product of the ISR correction and the detection
efficiency $(1+\delta)\cdot\epsilon$ is from two parts. One part is due to the statistical
uncertainty of the input cross section line shape, which is
estimated by varying the central value of the nominal input line shape
within $\pm$ 1$\sigma$ of the statistical uncertainty. The
$(1+\delta)\cdot\epsilon$ values for
each CM energy point are then recalculated. This process is repeated 100
times, and a Gaussian function is used to fit the
$(1+\delta)\cdot\epsilon$ distribution. The width of the Gaussian function
is taken as the corresponding systematic uncertainty. The other
uncertainty arises from the resonance parameters which are fixed in
the fit to the input cross section. The uncertainty from the
line-shape description is estimated with an alternative input cross
section line shape based on one resonance plus a power-law
function. The $(1+\delta)\cdot\epsilon$ value for each of CM energy
points is then recalculated, and the largest change is taken as the
systematic uncertainty. The total systematic uncertainty for the input
line shape is 2.4\% by adding both contributions in
quadrature.

\subsection{Total systematic uncertainty}

\noindent Assuming all sources are independent, the total systematic
uncertainty on the cross section measurement of 6.4\% is determined by
adding these sources in quadrature.

\section{Fit to the dressed cross section}

\noindent The potential resonances are searched for by analyzing the
dressed cross section ($\sigma^{\rm dressed}
=\frac{\sigma^{B}}{|1-\Pi|^2}$) of the $e^+e^-\to\Xi^0\bar{\Xi}^0$
reaction using the least $\chi^{2}$ method $\chi^{2} = \Delta
X^{T}V^{-1}\Delta X$.  Here $V$ is the covariance matrix incorporating
both the correlated and uncorrelated uncertainties among different CM
energy points, and $\Delta X$ is the vector of residuals between the
measured and the fitted cross sections. The diagonal elements of $V$
are the sum of the statistical uncertainties and uncorrelated
systematic uncertainties, added in quadrature. The off-diagonal elements of $V$
are correlated systematic uncertainties including the luminosity,
$\Xi^{0}$ reconstruction and branching fraction by assuming them to be
fully correlated for each CM energy.

\begin{figure}[!hbpt]
	\begin{center}
        \includegraphics[width=0.46\textwidth]{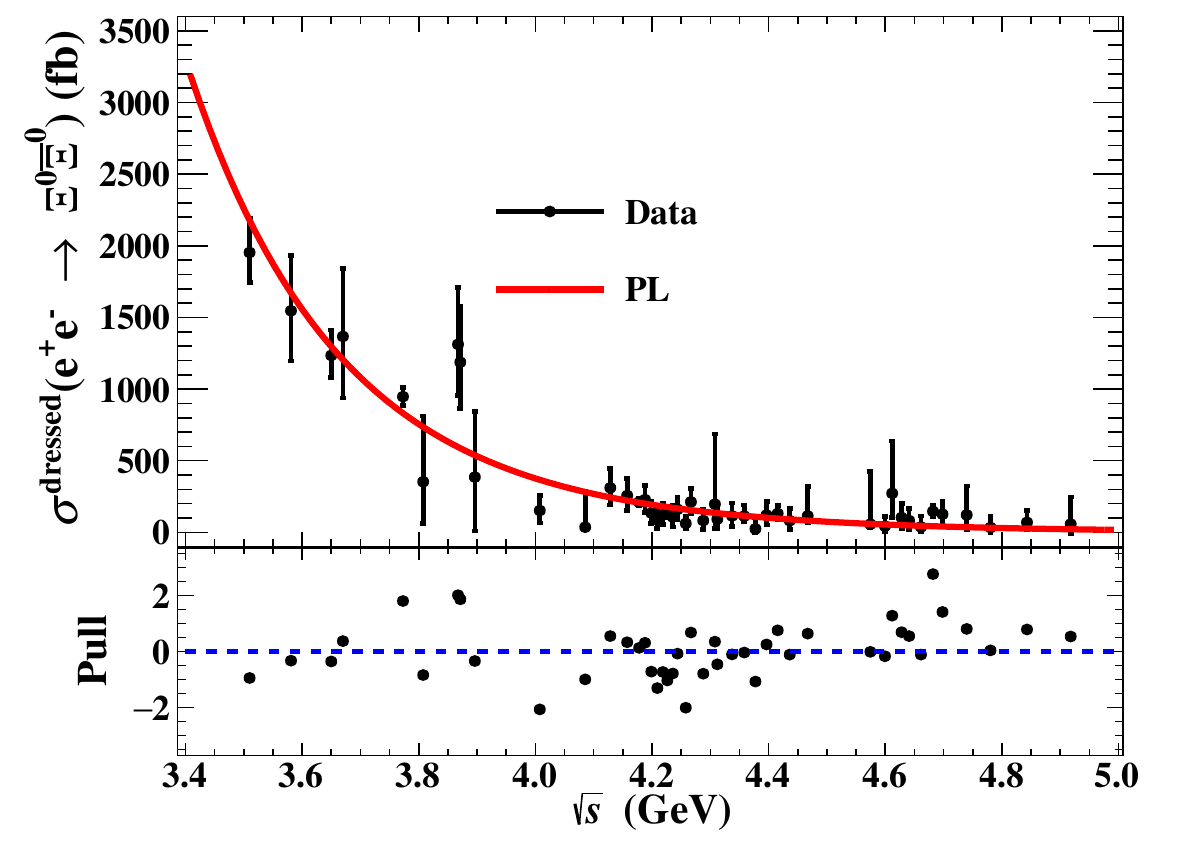}
        \includegraphics[width=0.46\textwidth]{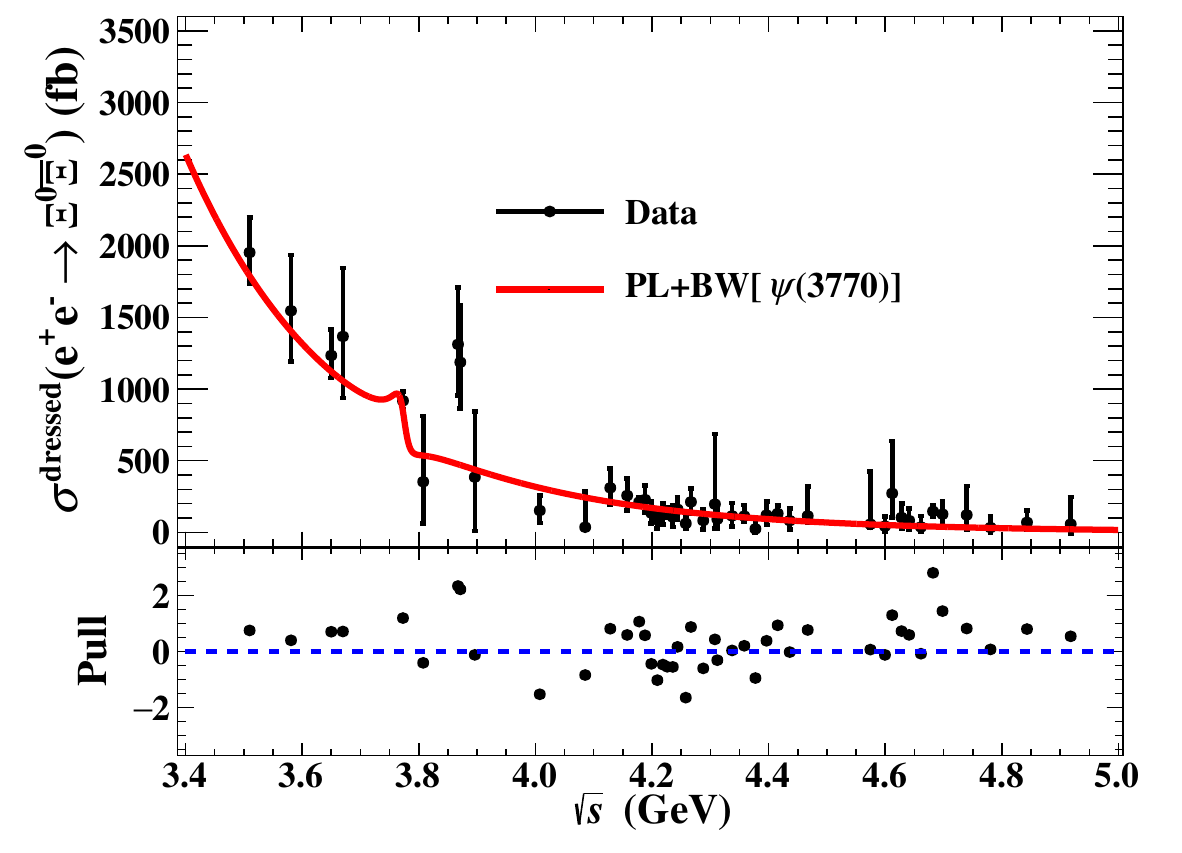}\\
        \includegraphics[width=0.46\textwidth]{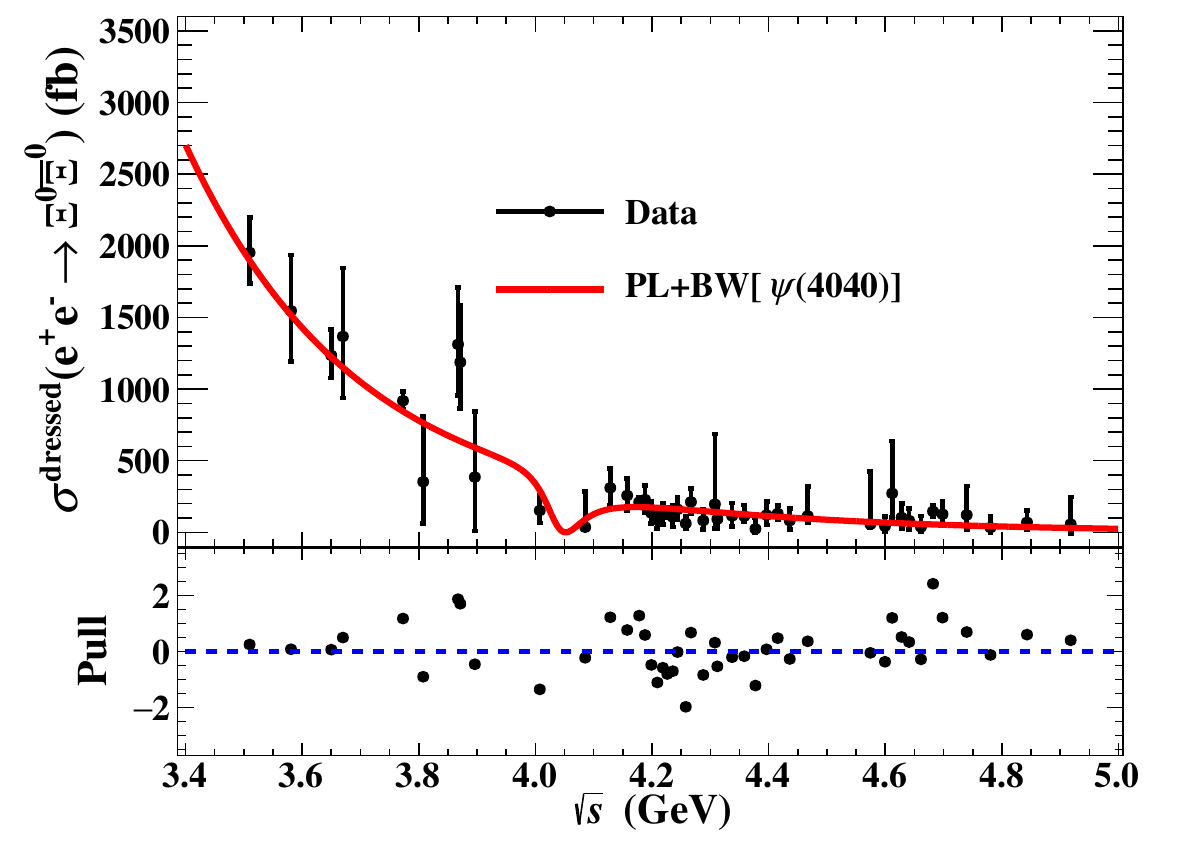}
        \includegraphics[width=0.46\textwidth]{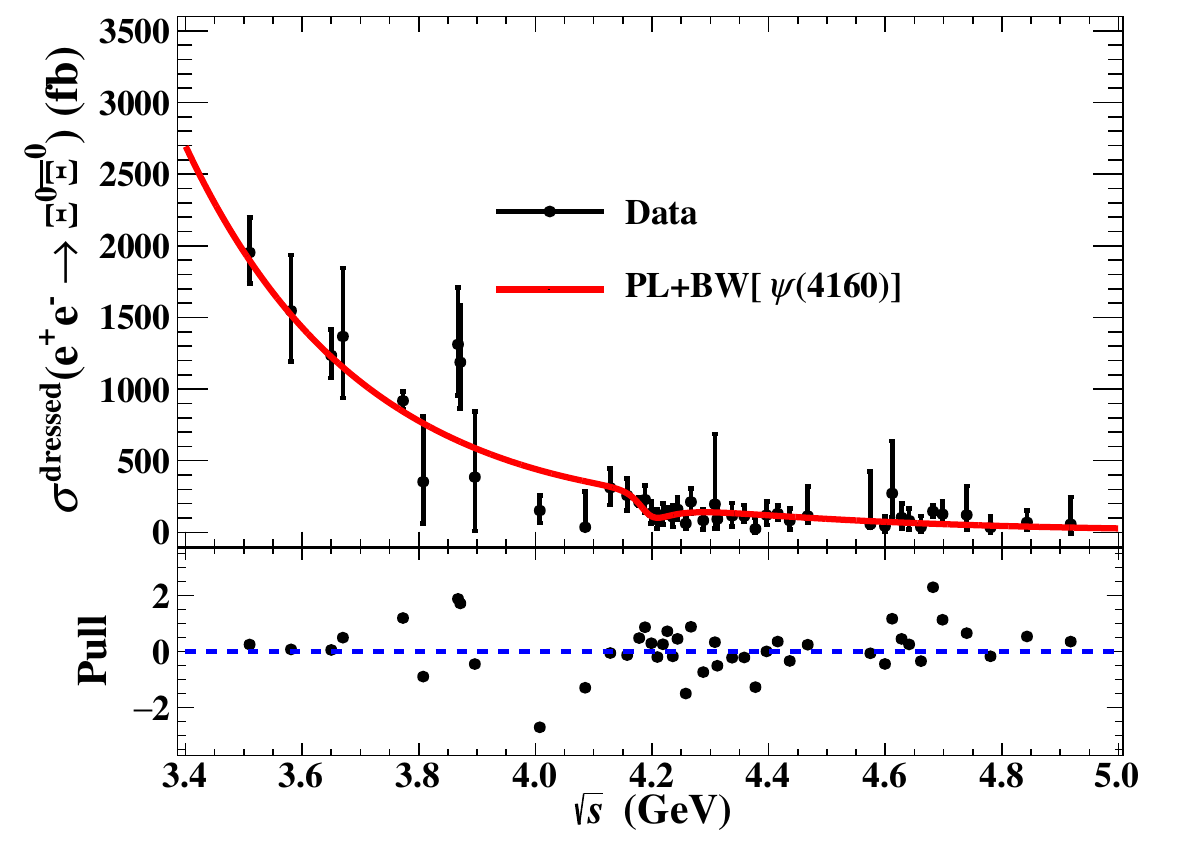}\\
        \includegraphics[width=0.46\textwidth]{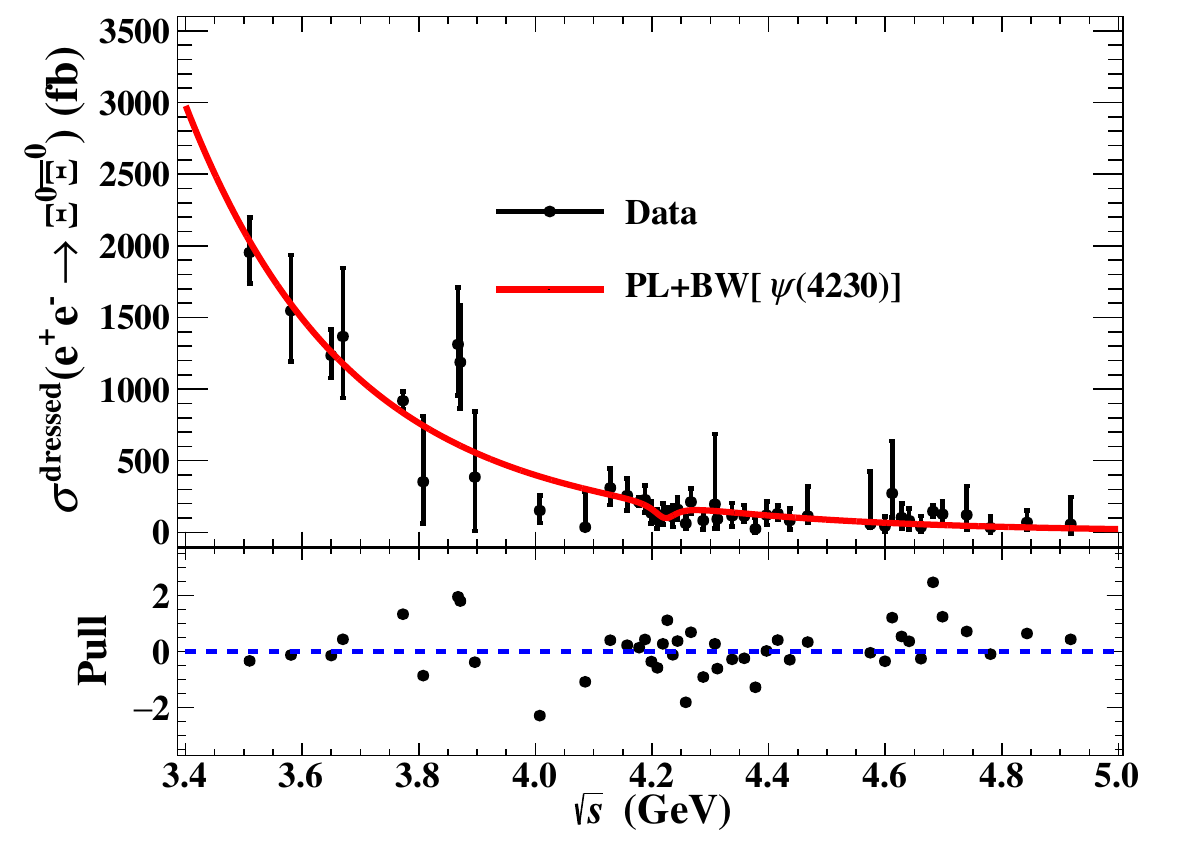}
        \includegraphics[width=0.46\textwidth]{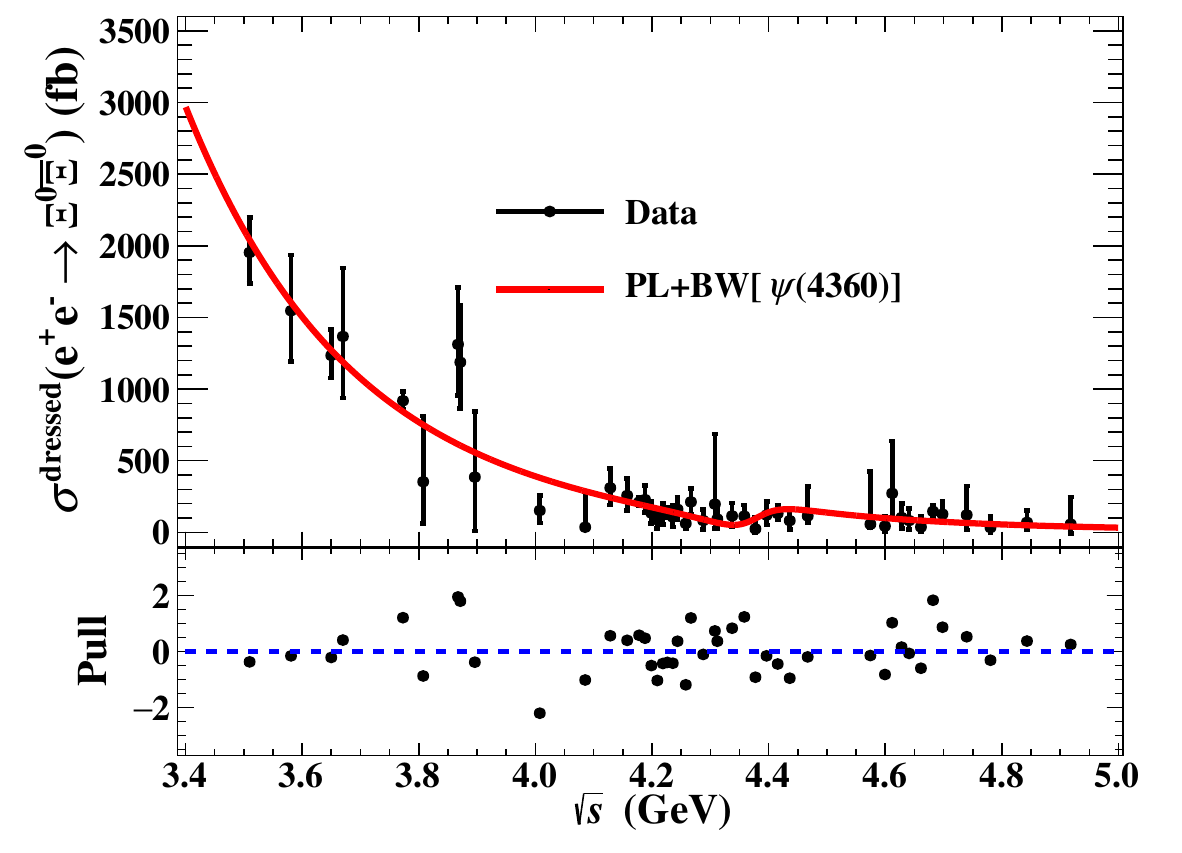}\\
        \includegraphics[width=0.46\textwidth]{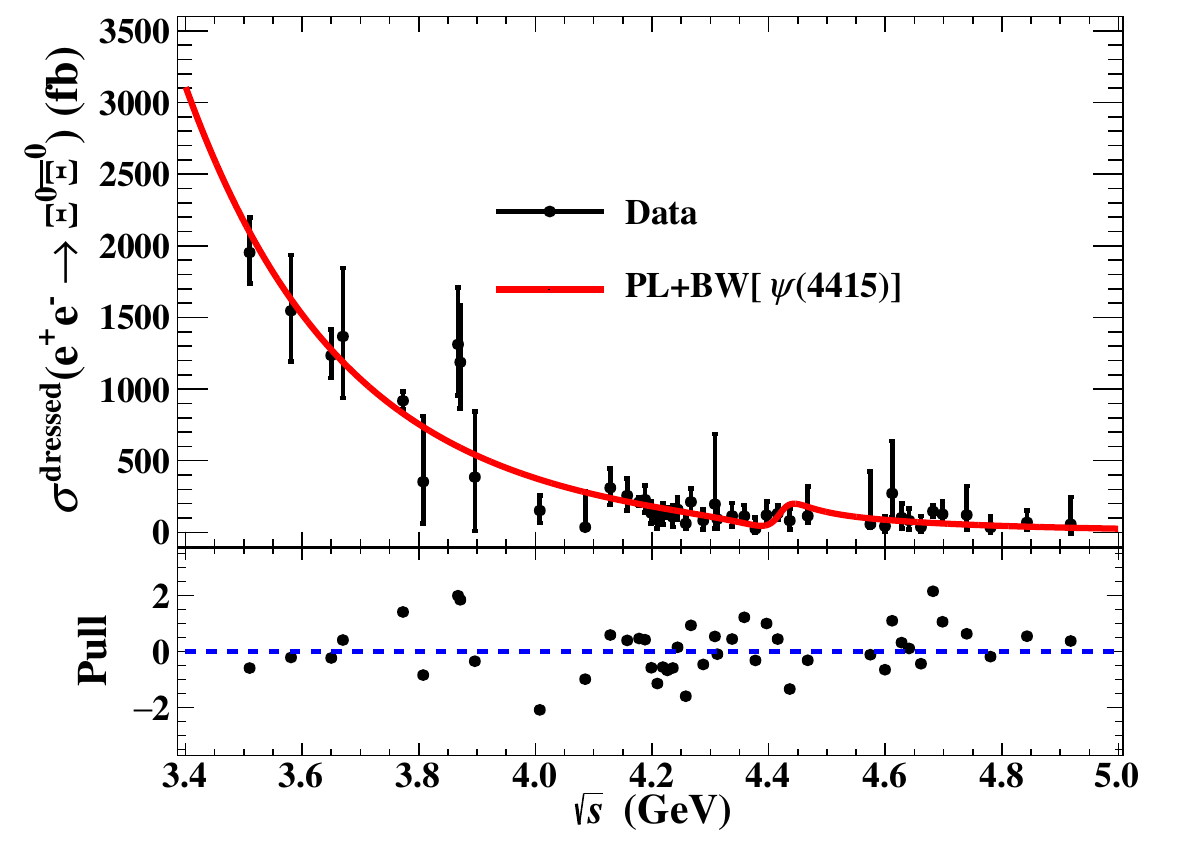}
        \includegraphics[width=0.46\textwidth]{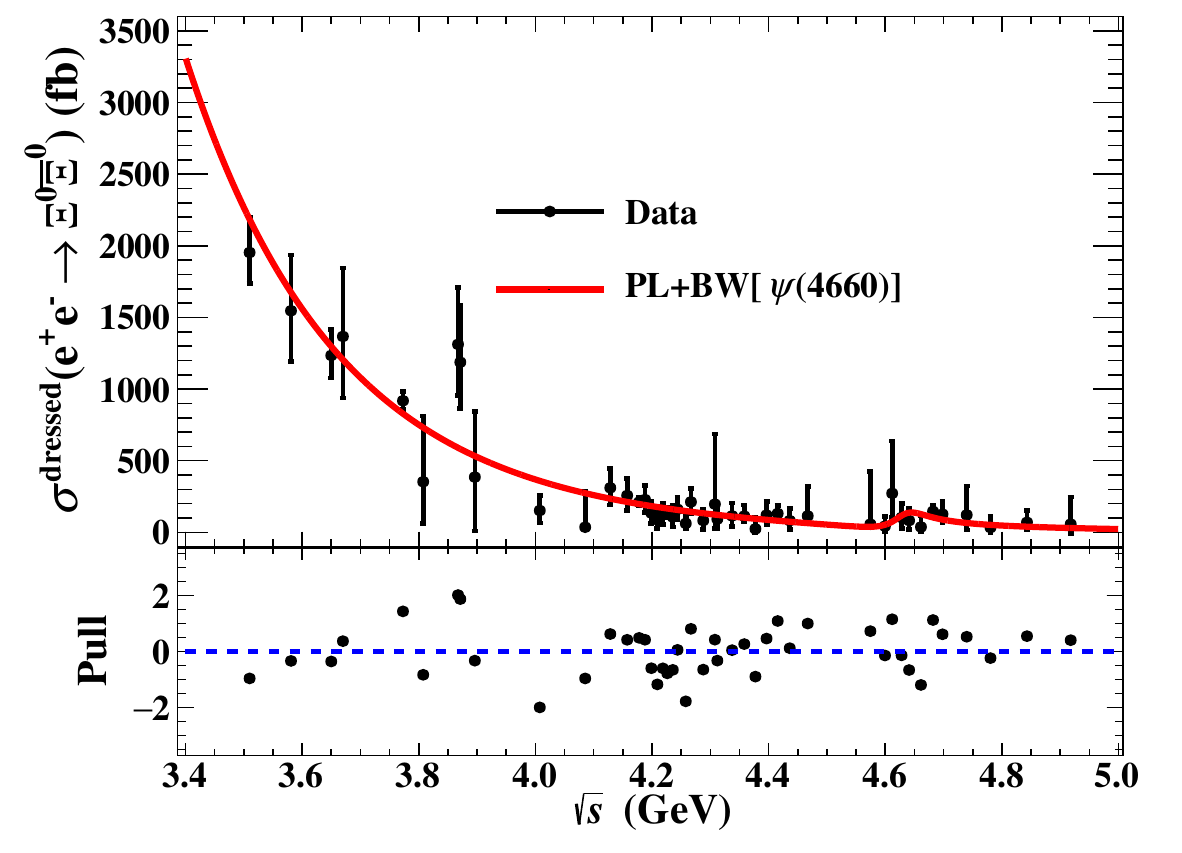}\\	
	\end{center}
	\caption{Fits to the dressed cross sections at CM energies
	from
	3.51 to \SI{4.95}{GeV} under the
        assumption of a PL function only (top left), or a PL function
        plus a resonance [i.e. $\psi(3770)$,
        $\psi(4040)$, $\psi(4160)$, $\psi(4230)$, $\psi(4360)$,
        $\psi(4415)$, or $\psi(4660)$]. Dots with error bars are
        the dressed cross sections, and the solid lines shows the
        fit results. The error bars include the statistical and
        systematic uncertainties summed in quadrature.}
	\label{Fig:XiXi::CS::Line-shape-3773}
\end{figure}

\begin{figure}[!htbp]
	\begin{center}
	\includegraphics[width=0.5\textwidth]{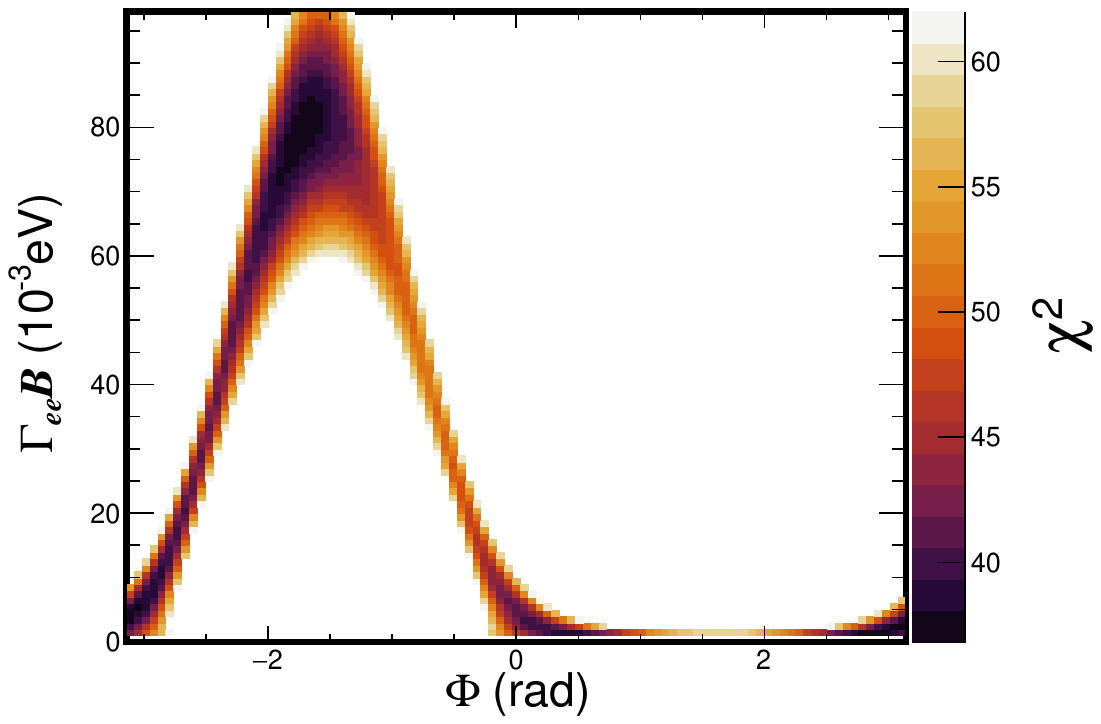}
    \put(-75, 75){\bf (a)}
    \includegraphics[width=0.5\textwidth]{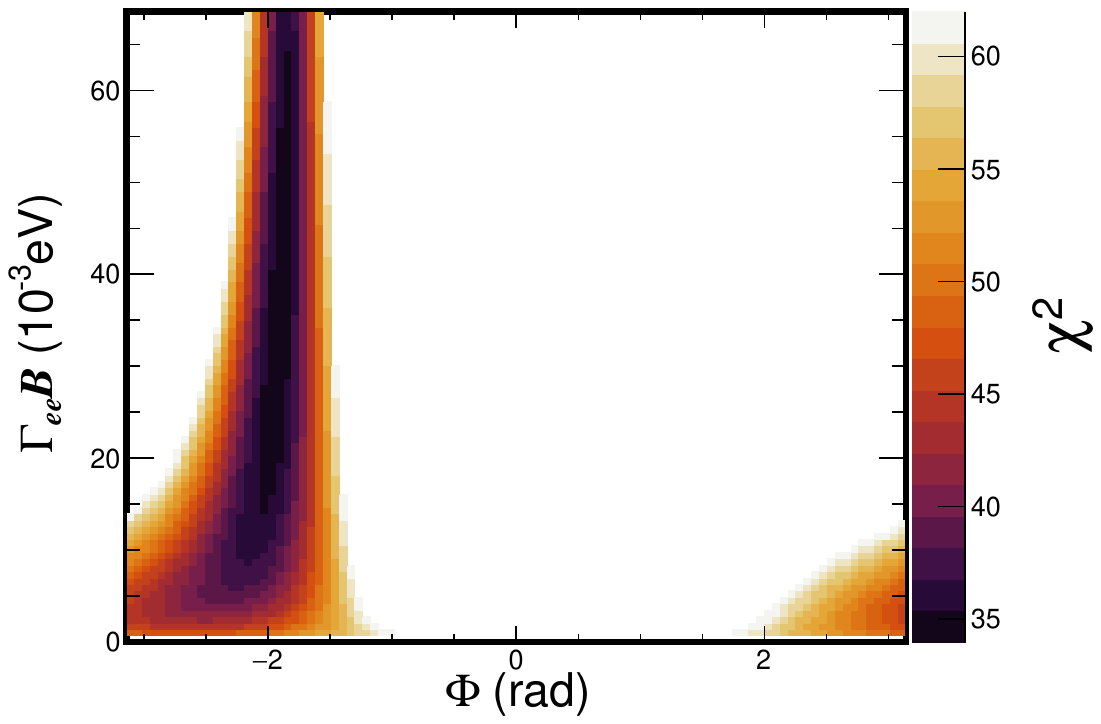}
    \put(-75, 75){\bf (b)}\\
    \includegraphics[width=0.5\textwidth]{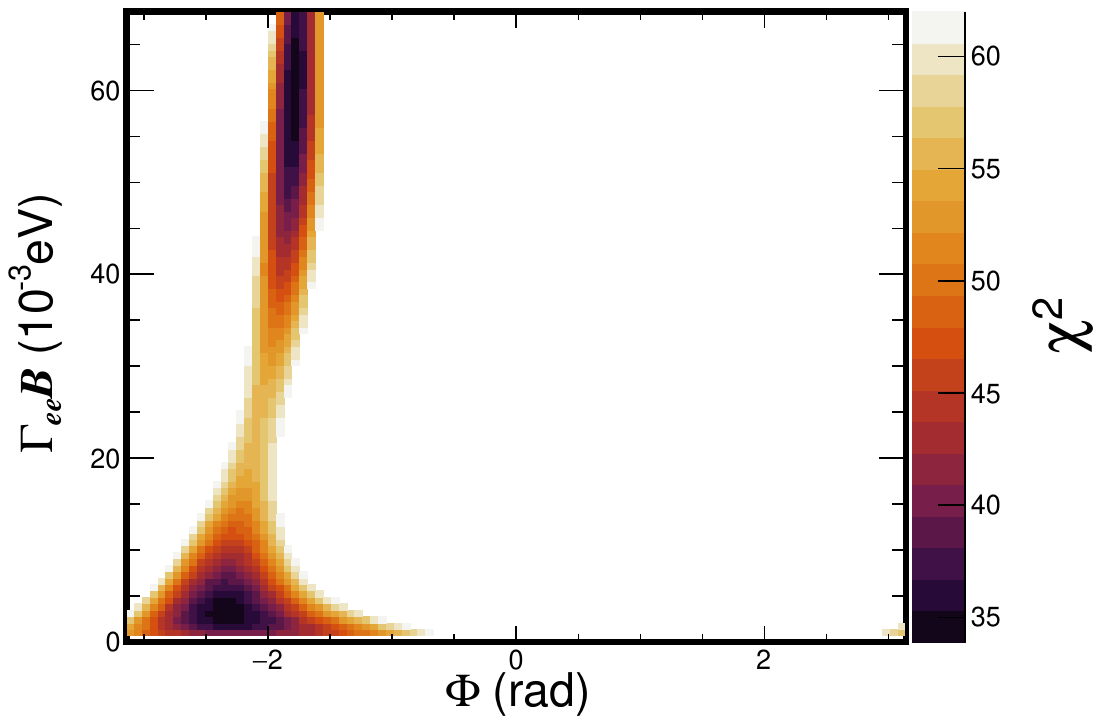}
    \put(-75, 75){\bf (c)}
    \includegraphics[width=0.5\textwidth]{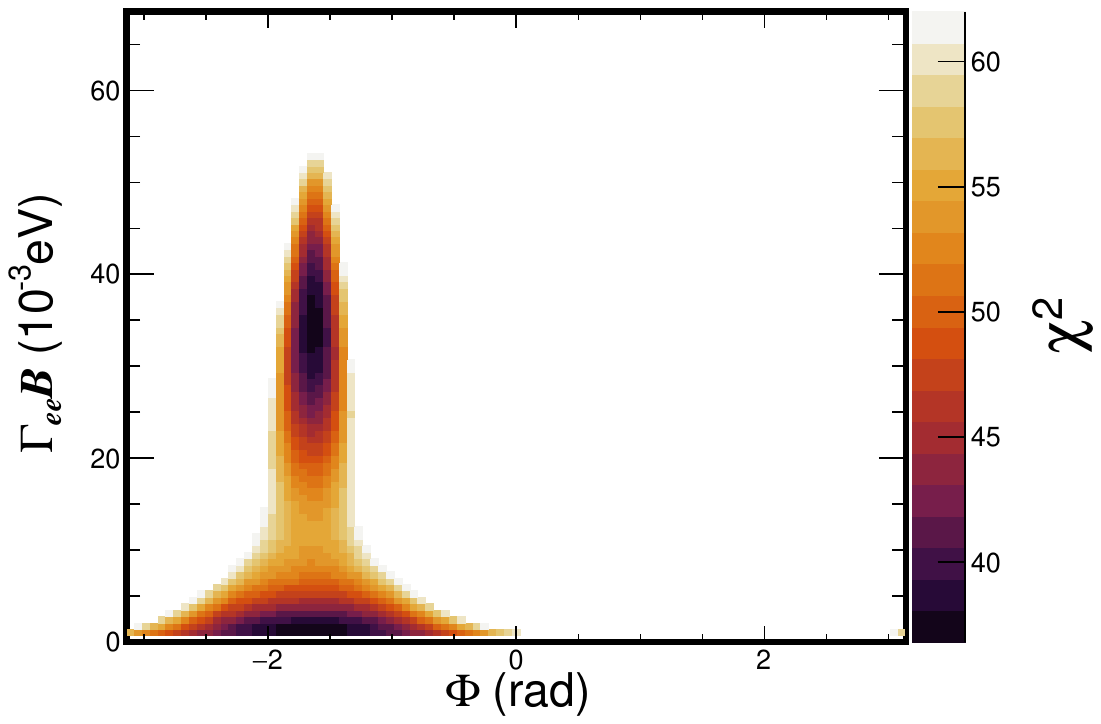}
    \put(-75, 75){\bf (d)}\\
    \includegraphics[width=0.5\textwidth]{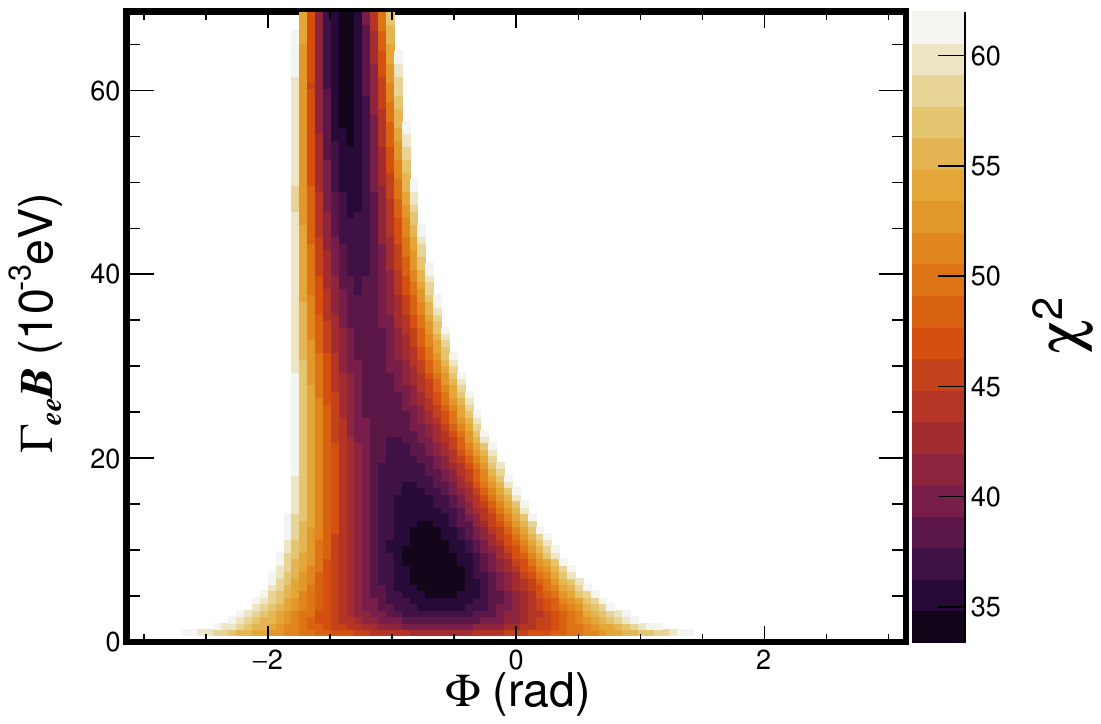}
    \put(-75, 75){\bf (e)}    
    \includegraphics[width=0.5\textwidth]{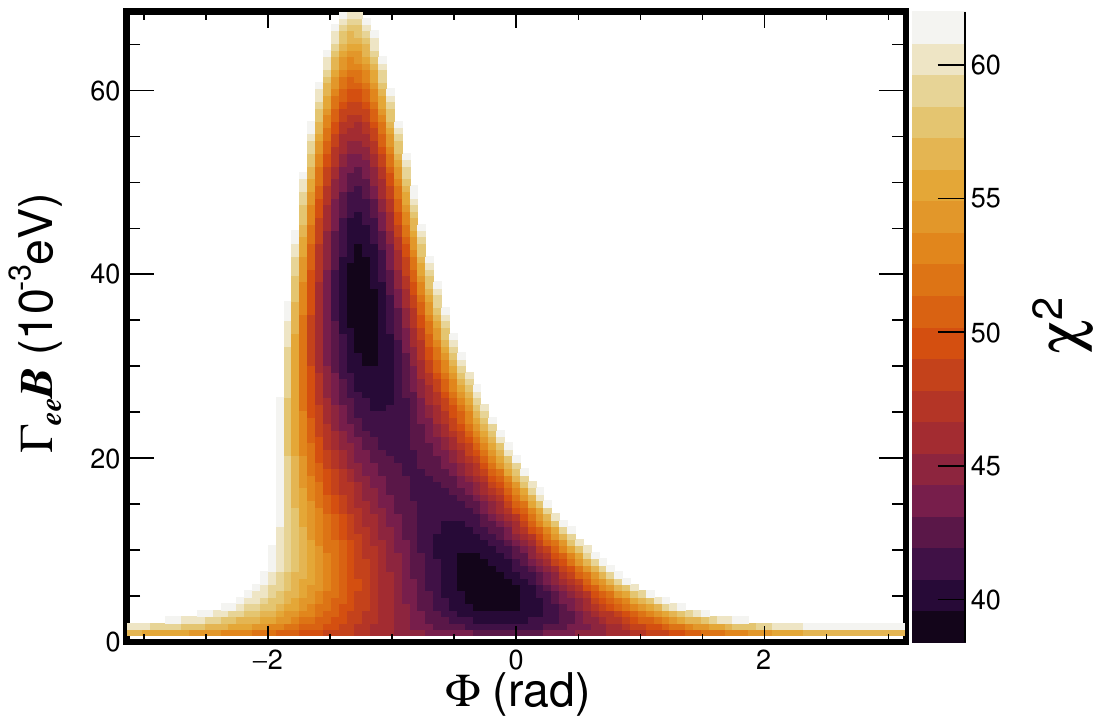}
    \put(-75, 75){\bf (f)}\\
    \includegraphics[width=0.5\textwidth]{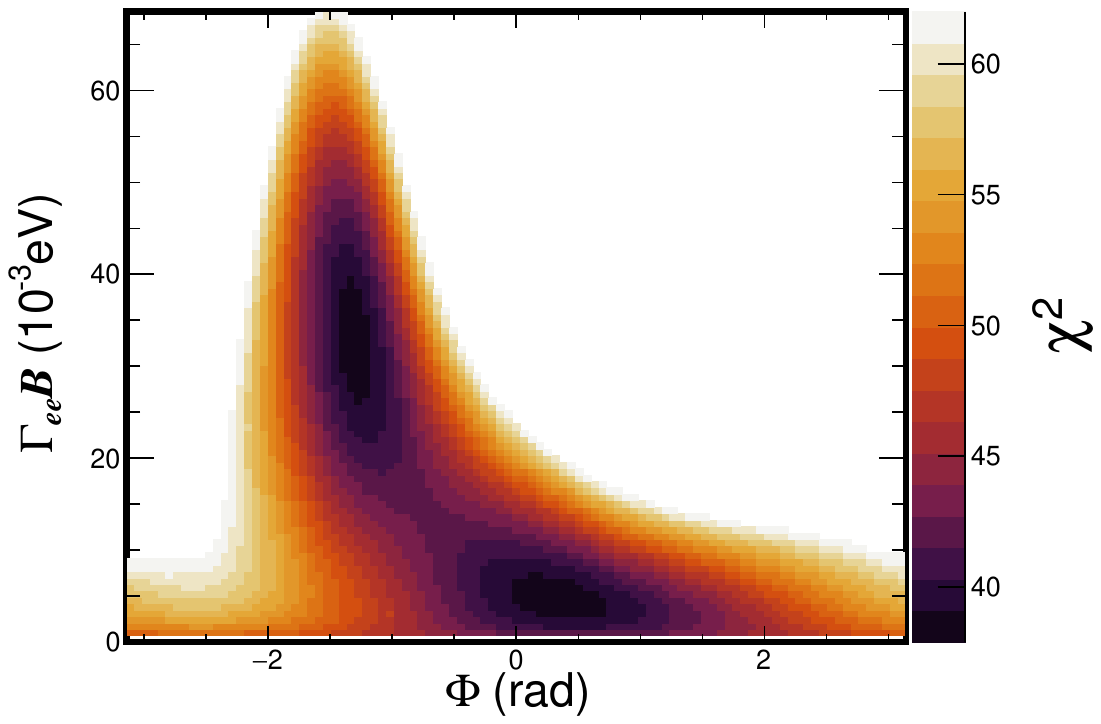}
    \put(-75, 75){\bf (g)}
	\end{center}
	\caption{Contour distributions of $\chi^2$ values in the
	$\Gamma_{ee} B-\Phi$ plane for (a) $\psi(3770)$, (b)
	$\psi(4040)$, (c) $\psi(4160)$, (d) $Y(4230)$, (e) $Y(4360)$,
	(f) $\psi(4415)$, and (g) $Y(4660)$ decaying into the
	$\Xi^0\bar{\Xi}^0$ final states. }
	\label{Fig:2D_Scan}
\end {figure}

Assuming the line shape of the dressed cross section of the
$e^+e^-\to\Xi^0\bar{\Xi}^0$ reaction includes a charmonium(-like)
amplitude, i.e., $\psi(3770)$, $\psi(4040)$, $\psi(4160)$,
$\psi(4230)$, $\psi(4360)$, $\psi(4415)$ or $\psi(4660)$, plus a
continuum contribution, a fit with the coherent sum of a power law(PL)
function plus a Breit-Wigner (BW) function \begin{equation}
    	\sigma^{\rm dressed}(\sqrt{s})=\left| c_{0}\frac{\sqrt{P(\sqrt{s})}}{(\sqrt{s})^{n}} +e^{i\phi}BW(\sqrt{s})\sqrt{\frac{P(\sqrt{s})}{P(M)}} \right|^2,
\label{eq:power_law}
\end{equation}
is used. 
Here $\phi$ is the interference angle between the BW function
\begin{equation}
    {\rm BW}(\sqrt{s}) =\frac{\sqrt{12\pi\Gamma_{ee}{\cal{B}}\Gamma}}{s-M^{2}+iM\Gamma},\label{eq:BW}
\end{equation} 
and the PL function, $c_0$ and $n$ are free parameters, $P(\sqrt{s})$
is the two-body PHSP factor ($P(\sqrt{s}) =
\frac{\sqrt{(s-(m_{\Xi^0}+m_{\bar\Xi^0})^2)(s-(m_{\Xi^0}-m_{{\bar\Xi}^0})^2)}}{2\sqrt{s}}$),
$M$ and $\Gamma$ are the resonance mass and total width, respectively,
fixed according to the PDG values~\cite{PDG2022}.
$\Gamma_{ee}{\cal{B}}$ is the product of the electronic partial width
and the branching fraction for each assumed charmonium(-like) state
decaying into the $\Xi^0\bar{\Xi}^0$ final state.
Figure~\ref{Fig:XiXi::CS::Line-shape-3773} shows fits to the dressed
cross section by eq.~(\ref{eq:power_law}) with a PL function only and
with a single charmonium(-like) amplitude, fitted one at a time. The
parameters with the PL function only are fitted to be $c_0 = 1.1 \pm
0.3, n = 7.7 \pm 0.2$. The parameters with the assumed
charmonium(-like) amplitude combined with their multiple solutions are
summarized in table~\ref{tab:multisolution}, where the possible
multiple solutions are evaluated based on a two-dimensional scan
method which scans all the pairs of $\Gamma_{ee}B$ and $\phi$ in
parameter space as shown in figure~\ref{Fig:2D_Scan}.  The
significance for each resonance considering the systematic uncertainty
is calculated by comparing the change of $\chi^{2}/\rm{n.d.f}$ with
and without including the resonance in the fit, where $\rm{n.d.f}$
represents the number of degrees of freedom, and no significant
resonance is found.  Consequently, $\Gamma_{ee}{\cal{B}}$ and its
upper limit at the 90\% C.L. using a Bayesian approach
\cite{Zhu:2008ca} and including the systematic uncertainty for each
assumed charmonium(-like) state decaying into the $\Xi^0\bar{\Xi}^0$
final state are listed in table~\ref{tab:multisolution}.
\begin{table}[h]
    \caption{The fitted resonance parameters for
    $\Gamma_{ee}\mathcal{B}$ $(10^{-3}~\rm{eV})$ and $\phi$ (rad) with two solutions. The fit procedure includes both statistical and systematic uncertainties. Here $\chi^2/\rm{n.d.f}$
     indicates the fit quality for each assumed resonance. }
    \centering
    \begin{tabular}{l r@{  }l r@{  }l l l}
        \hline
        \hline
        Parameter                                              &\multicolumn{2}{c}{Solution I}         &\multicolumn{2}{c}{Solution II}       &$\chi^2/\rm{n.d.f}$
        \\
        \hline
$\phi_{\psi(3770)}$              &$-1.7$&$\pm$ 0.2                  &$2.9$&$\pm$ 0.8         \\ 
$\Gamma_{ee}\mathcal{B}_{\psi(3770)}$   &$79.1$&$\pm$ 6.5 $(<89.0)$  &$1.1$&$\pm$ 3.3         & $37/(45-4)$&\\

$\phi_{\psi(4040)}$             &$-1.9 $&$\pm$ 0.1                    &$-$ \\
$\Gamma_{ee}\mathcal{B}_{\psi(4040)}$ &$34.2  $&$\pm$ 19.0 $(<83.4)$ &$-$ & &$34/(45-4)$&\\

$\phi_{\psi(4160)}$              &$-1.8 $&$\pm$ 0.1                  &$-2.3 $&$\pm$ 0.1  \\
$\Gamma_{ee}\mathcal{B}_{\psi(4160)}$    &$59.2 $ &$\pm$ 3.8 $(<69.0)$   &$3.0  $&$\pm$ 1.1   &$34/(45-4)$&\\

$\phi_{Y(4230)}$             &$-1.6 $&$\pm$ 0.1                     &$-1.6 $&$\pm$ 0.2  \\
$\Gamma_{ee}\mathcal{B}_{Y(4230)}$  &$34.0 $&$\pm$ 2.7 $(<40.4)$    &$1.0  $&$\pm$ 0.5   &$37/(45-4)$&\\

$\phi_{Y(4360)}$            &$-1.4 $&$\pm$ 0.1                      &$-0.7 $&$\pm$ 0.2   \\
$\Gamma_{ee}\mathcal{B}_{Y(4360)}$ &$66.5 $&$\pm$ 8.1 $(<84.5)$     &$7.9  $&$\pm$ 3.4    &$34/(45-4)$&\\

$\phi_{\psi(4415)}$         &$-1.2 $&$\pm$ 0.1                      &$-0.2 $&$\pm$ 0.2  \\
$\Gamma_{ee}\mathcal{B}_{\psi(4415)}$   &$36.3 $&$\pm$ 5.0 $(<48.0)$    &$5.6  $&$\pm$ 3.0  &$39/(45-4)$&\\

$\phi_{Y(4660)}$            &$-1.3 $&$\pm$ 0.1                       &$0.3 $&$\pm$ 0.3  \\
$\Gamma_{ee}\mathcal{B}_{Y(4660)}$ &$32.8 $&$\pm$ 5.4 $(<45.0)$     &$4.5  $&$\pm$ 1.9   &$38/(45-4)$&\\
        \hline
        \hline
    \end{tabular}
    \label{tab:multisolution}
\end{table}

\section{Summary}

\noindent In summary, using a total integrated luminosity of 30
fb$^{-1}$ of $e^+e^-$ collision data collected by the BESIII detector
at BEPCII for CM energies between 3.51 and 4.95 GeV, we measure the
Born cross sections and the effective form factors at forty-five CM
energy points for the $e^+e^-\ar\Xi^0\bar{\Xi}^0$ reaction.  The dressed cross section of this reaction is fitted under the assumption of a single charmonium(-like) amplitude
plus a continuum contribution. No obvious signal for any assumed
charmonium(-like) state [i.e., $\psi(3770)$, $\psi(4040)$,
$\psi(4160)$, $\psi(4230)$, $\psi(4360)$, $\psi(4415)$, or
$\psi(4660)$] is found.  The products of the branching fraction and
electronic partial width for each assumed charmonium(-like) state
decaying into the $\Xi^0\bar{\Xi}^0$ final state as well as the upper
limits at the 90\% C.L. are provided, which can be used to further
constrain theoretical models~\cite{Close:2005iz, Qian:2021neg}.  In
addition, the ratios of the Born cross section and the effective form
factor between this work and $e^+e^-\to\Xi^+\bar\Xi^-$, as shown in
figure~\ref{Fig:ratio_of_sig}, can be used to test isospin symmetry
and the vector meson dominance
model~\cite{Iachello:1972nu,Iachello:2004aq,Bijker:2004yu,
Yang:2019mzq,Li:2021lvs,Dai:2023vsw}.  The results of this study
provide important experimental information on the correlation between
vector charmonium(-like) states and the $e^+e^-\to\Xi^0\bar\Xi^0$
production.

\newpage
\acknowledgments
\noindent The BESIII Collaboration thanks the staff of BEPCII and the IHEP computing center for their strong support. This work is supported in part by National Key R\&D Program of China under Contracts Nos. 2023YFA1606000, 2020YFA0406300, 2020YFA0406400; National Natural Science Foundation of China (NSFC) under Contracts Nos. 
12075107, 12247101, 11635010, 11735014, 11935015, 11935016, 11935018, 12025502, 12035009, 12035013, 12061131003, 12192260, 12192261, 12192262, 12192263, 12192264, 12192265, 12221005, 12225509, 12235017, 12361141819;
the 111 Project under Grant No. B20063; 
 the Chinese Academy of Sciences (CAS) Large-Scale Scientific Facility Program; the CAS Center for Excellence in Particle Physics (CCEPP); Joint Large-Scale Scientific Facility Funds of the NSFC and CAS under Contract No. U1832207; 100 Talents Program of CAS; The Institute of Nuclear and Particle Physics (INPAC) and Shanghai Key Laboratory for Particle Physics and Cosmology; German Research Foundation DFG under Contracts Nos. FOR5327, GRK 2149; Istituto Nazionale di Fisica Nucleare, Italy; Knut and Alice Wallenberg Foundation under Contracts Nos. 2021.0174, 2021.0299; Ministry of Development of Turkey under Contract No. DPT2006K-120470; National Research Foundation of Korea under Contract No. NRF-2022R1A2C1092335; National Science and Technology fund of Mongolia; National Science Research and Innovation Fund (NSRF) via the Program Management Unit for Human Resources \& Institutional Development, Research and Innovation of Thailand under Contracts Nos. B16F640076, B50G670107; Polish National Science Centre under Contract No. 2019/35/O/ST2/02907; Swedish Research Council under Contract No. 2019.04595; The Swedish Foundation for International Cooperation in Research and Higher Education under Contract No. CH2018-7756; U. S. Department of Energy under Contract No. DE-FG02-05ER41374


\newpage
{\bf \noindent The BESIII collaboration}\\
{\small
M.~Ablikim$^{1}$, M.~N.~Achasov$^{4,c}$, P.~Adlarson$^{76}$, O.~Afedulidis$^{3}$, X.~C.~Ai$^{81}$, R.~Aliberti$^{35}$, A.~Amoroso$^{75A,75C}$, Y.~Bai$^{57}$, O.~Bakina$^{36}$, I.~Balossino$^{29A}$
% ...
\\
\\
{\it
$^{1}$ Institute of High Energy Physics, Beijing 100049, People's Republic of China\\
% ...
$^{3}$ Bochum  Ruhr-University, D-44780 Bochum, Germany\\
$^{4}$ Budker Institute of Nuclear Physics SB RAS (BINP), Novosibirsk 630090, Russia\\
% ...
$^{29}$ INFN Sezione di Ferrara, (A)INFN Sezione di Ferrara, I-44122, Ferrara, Italy; (B)University of Ferrara,  I-44122, Ferrara, Italy\\
% ...
$^{35}$ Johannes Gutenberg University of Mainz, Johann-Joachim-Becher-Weg 45, D-55099 Mainz, Germany\\
$^{36}$ Joint Institute for Nuclear Research, 141980 Dubna, Moscow region, Russia\\
% ...
$^{57}$ Southeast University, Nanjing 211100, People's Republic of China\\
% ...
$^{75}$ University of Turin and INFN, (A)University of Turin, I-10125, Turin, Italy; (B)University of Eastern Piedmont, I-15121, Alessandria, Italy; (C)INFN, I-10125, Turin, Italy\\
$^{76}$ Uppsala University, Box 516, SE-75120 Uppsala, Sweden\\
% ...
$^{81}$ Zhengzhou University, Zhengzhou 450001, People's Republic of China\\
\vspace{0.2cm}
% ...
$^{c}$ Also at the Novosibirsk State University, Novosibirsk, 630090, Russia\\
% ...
}}
\end{document}